\documentclass[journal]{IEEEtran}
\usepackage{amsmath, amssymb, amsfonts}
\usepackage{bm}
\usepackage{algorithm}
\usepackage{algorithmic}
\usepackage{multirow}
\usepackage{booktabs}
\usepackage{graphicx}
\usepackage{cite}
\usepackage{float}
\usepackage{color}
\usepackage{tikz}
\usetikzlibrary{shapes.geometric, arrows.meta, positioning, calc, fit}
\usepackage{amsmath, bm}

\definecolor{uncertainpurple}{rgb}{0.50,0.00,0.50}

\def\argmin{\mathop{\arg\min}}

\let\OldIncludeGraphics\includegraphics
\renewcommand{\includegraphics}[2][]{
  \IfFileExists{#2}{\OldIncludeGraphics[#1]{#2}}{
    \fbox{\parbox[c][3cm][c]{0.90\linewidth}{\centering\footnotesize Missing figure\\Original figure file was not uploaded.}}
  }
}

\title{A Self-Generated Local Structural Prior for Resolving Target Merging in Electrical Impedance Tomography (EIT)}
\author{}
\author{Junkang~Liu and Yizhuang Song
\thanks{This work was supported by National Natural Science Foundation of China (Grant No. 12271312). (Corresponding author: Yizhuang Song.)}
\thanks{Junkang Liu and Yizhuang Song are with School of Mathematics and Statistics, Shandong Normal University, Jinan, Shandong, 250014, P. R. China. (e-mail: ysong@sdnu.edu.cn).}}

\date{}

\begin{document}

\maketitle

\begin{abstract}
Electrical impedance tomography (EIT) is an ill-posed inverse imaging technique with limited and spatially nonuniform resolution. Conductivity targets located in regions with degraded spatial sensitivity may be reconstructed as a single connected region, resulting in the loss of spatial separability between neighboring structures. In this work, we propose a self-generated local structural prior for suppressing target merging in EIT reconstruction. First, an initial conductivity image is obtained from the measured boundary data using an existing EIT reconstruction method. This reconstruction is subsequently used as a reference image to construct a Gaussian similarity matrix according to the intensity similarity between image pixels. The adaptively chosen eigenvector of the covariance matrix is then employed to extract the principal structural information of the reference image. A localized structural mask is constructed and incorporated into a subsequent EIT reconstruction to preserve the spatial separability of neighboring targets. Numerical and water tank experimental results demonstrate that the proposed method effectively separates neighboring conductivity targets that are merged by conventional EIT reconstruction methods.
\end{abstract}


\section{Introduction}

\IEEEPARstart{E}{lectrical} Impedance Tomography (EIT) is a medical imaging modality that visualizes the internal structure and functional properties of human organs by imaging the body's admittivity distribution. In EIT, a time-harmonic current $i(t) = I\sin(\omega t)$ is sequentially injected via a pair of electrodes, and the induced voltage differences are measured across the remaining electrode pairs \cite{Cheney1999}. Here, $\omega$ denotes the angular frequency of the injected current, which satisfies $\omega/2\pi \leq $1 MHz for EIT systems, and $I$ represents the current amplitude. To reconstruct the admittivity distribution, EIT solves the inverse problem using prior knowledge including the body geometry, electrodes configurations, the current injection patterns, and the measured voltage differences. 

EIT offers several advantages, including portability, non-invasiveness, low cost, and long-term stable high-speed data acquisition, and therefore holds substantial potential as a medical device for clinical applications, particularly in intensive care units (ICUs). However, local variations in internal admittivity induce only minor changes in boundary-measured voltage differences, which renders the inverse problem severely ill-posed. Furthermore, such ill-posedness greatly limits the practical deployment of EIT in real-world scenarios.

Numerous regularization strategies have been proposed to address the ill-posedness of the EIT inverse problem. These strategies can be broadly categorized into model-driven and data-driven regularizers.

For model-driven regularizers, Tikhonov regularization is a representative classical scheme \cite{Cheney1999,Vauhkonen1998}. Nevertheless, due to its inherent smoothness prior, this method tends to over-smooth sharp anatomical edges of internal organs. Multiplicative regularization \cite{Zhang2019} avoids the manual selection of regularization parameters. However, since it adopts an $\ell_2$-norm-based regularizer, it still suffers from edge blurring artifacts. Total variation (TV) and its variants have been widely adopted to preserve structural edges during reconstruction \cite{Rudin1992,Borsic2010,Li2019,Shi2021a,Shi2021b}. Despite their edge-preserving capability, these methods typically produce blocky artifacts and involve intensive computational cost, as the primal-dual algorithm is generally required to address the non-differentiable optimization problem \cite{Borsic2010}. 

To improve computational efficiency, anisotropic TV regularization was developed with the alternating direction method of multipliers (ADMM) for accelerated iteration \cite{Gonzalez2017}. However, this approach may introduce axis-aligned geometric distortions in reconstructed images. Furthermore, Lee {\it et al.} \cite{Lee2018} proposed fidelity-embedded regularization to achieve robust conductivity reconstruction without parameter tuning. Nevertheless, this method inevitably compromises spatial resolution and may lead to lung merging artifacts for obese subjects \cite{Seo2019}. To simultaneously accelerate reconstruction and enhance edge preservation, Song {\it et al.} \cite{Song2022} presented a nonlinear weighted anisotropic total variation regularization strategy, which improves the separation performance of bilateral lung regions for obese cases.

Recently, data-driven regularization techniques have emerged with the rapid development of deep learning. A variational autoencoder (VAE)-based manifold learning method was proposed in \cite{Seo2019}, which adopts the reconstruction results from \cite{Lee2018} as training data to realize dimensionality reduction during image reconstruction. This method gives superior reconstruction performance over the approach in \cite{Lee2018} for obese subjects, particularly in achieving accurate separation of the bilateral lungs. However, the difficulty in acquiring sufficient labeled training data limits the practical application of supervised learning schemes, thereby necessitating unsupervised regularization strategies. To address this issue, Liu \textit{et al.} \cite{Liu2023DeepEIT,Xia2024} exploited the deep image prior (DIP) framework, where a U-Net architecture is employed to serve as an implicit regularizer. Nevertheless, this approach is sensitive to the number of iterations, and early stopping is essentially required to avoid overfitting and guarantee stable reconstruction quality. For a more comprehensive review and in-depth discussion of recently developed deep-learning-based EIT reconstruction methods, we refer readers to the work of Denker \textit{et al.} \cite{Denker2025} and the references therein.
Despite substantial progress in data-driven regularization, such approaches still face challenges in balancing reconstruction accuracy, computational efficiency, robustness, and explainability. 

To address the above issue, we propose a self-generated local structural prior for EIT reconstruction. Unlike conventional structural-prior approaches that directly introduce anatomical information from external imaging modalities, the proposed method first obtains an initial EIT reconstruction and uses it as a reference image for extracting structural information. A Gaussian similarity covariance matrix is constructed according to the intensity similarity between pixels in the reference image. 

The adaptively chosen eigenvector from its leading eigenvectors is used to construct a localized structural mask to prevent target merging. The main contributions of this work are threefold: (1) a self-generated reference-image strategy is developed to obtain structural prior information directly from an initial EIT reconstruction without requiring additional anatomical imaging data; (2) a Gaussian-kernel eigenvector representation is introduced to extract dominant structural information from the reference image; and (3) a localized eigenvector-based structural mask is proposed to specifically suppress target merging in spatially ambiguous regions.

Numerical and water tank experimental results validate the advantages in separating neighboring conductivity targets that are merged by conventional EIT reconstruction methods.

\section{Methods}
\subsection{Background of EIT}

Suppose that the object to be imaged occupies a domain $\Omega \subset \mathbb{R}^n$ ($n=2,3$) with a smooth boundary
$\partial\Omega$. The admittivity distribution in $\Omega$ is given by $\gamma = \sigma + \mathrm{i}\omega\epsilon$,
where $\mathrm{i}=\sqrt{-1}$, and $\sigma$ and $\epsilon$ denote, respectively, the electrical conductivity and permittivity distributions in $\Omega$ at the angular frequency $\omega$. In EIT, a sequence of time-harmonic currents, $i(t)=I\sin(\omega t)$, is injected successively through adjacent electrode pairs $(\mathcal{E}_j,\mathcal{E}_{j+1})$. In the low-frequency regime considered here, $0<\omega/(2\pi)\leq 1\,\mathrm{MHz}$, the imaginary parts of the admittivity and the resulting electric potential are assumed to be negligible compared with their respective real parts. Consequently, for the $j$-th current injection, the electric potential $u^j$ is approximately governed by the following elliptic partial differential equation (PDE) with mixed boundary conditions:
\begin{equation}\label{eq:forward}
\left\{
\begin{split}
    & \nabla \cdot (\sigma \nabla u^j) = 0, \quad \text{in } \Omega \\
    & \sigma \frac{\partial u^j}{\partial {\bf n}} = 0, \quad \text{on } \partial\Omega \setminus \overline{\cup_{i=1}^E \mathcal{E}_i} \\
    & u^j + z_i \sigma \frac{\partial u^j}{\partial {\bf n}}\Big|_{\mathcal{E}_i} = U_i^j[\sigma], \quad i = 1, 2, \dots, E \\
    & \int_{\mathcal{E}_j} \sigma \frac{\partial u^j}{\partial {\bf n}} dS = I = -\int_{\mathcal{E}_{j+1}} \sigma \frac{\partial u^j}{\partial {\bf n}} dS \\
    & \int_{\mathcal{E}_k} \sigma \frac{\partial u^j}{\partial {\bf n}} dS = 0, \quad k \in \{1, 2, \dots, E\} \setminus \{j, j+1\}
\end{split}
\right.
\end{equation}
Here, the electrode indices are interpreted cyclically, such that $\mathcal{E}_{E+1}=\mathcal{E}_1$. Moreover, $z_i$ denotes the contact impedance between the $i$-th electrode $\mathcal{E}_i$ and the boundary $\partial\Omega$, $\mathbf{n}$ is the outward unit normal vector on $\partial\Omega$, $\sigma$ is the electrical conductivity, and $U_i^j[\sigma]$ denotes the voltage measured on $\mathcal{E}_i$ under the $j$-th current-injection pattern.

To ensure the uniqueness of the electric potential, the grounding condition, $\sum_{i=1}^{E}U_i^j[\sigma]=0$, is imposed.  For each current-injection pattern $j$, the voltage differences $V_i^j[\sigma] := U_i^j[\sigma]-U_{i+1}^j[\sigma]$ are measured across adjacent electrode pairs $(\mathcal{E}_i,\mathcal{E}_{i+1})$, where the electrode indices are interpreted cyclically. To reduce the influence of contact impedance, voltage measurements involving either of the current-carrying electrodes, $\mathcal{E}_j$ and $\mathcal{E}_{j+1}$, are excluded. Thus, for each current-injection pattern, voltage measurements from $E-3$ electrode pairs are retained. Consequently, an $E$ channel EIT system provides a total of $E(E-3)$ voltage-difference measurements for image reconstruction.

The inverse problem of EIT aims to reconstruct the conductivity distribution $\sigma$ from the measured voltage differences $\{V_i^j\}$ using the reciprocity relation
\begin{equation}\label{eq:reciprocity}
    \frac{1}{I}\int_{\Omega} \sigma(\mathbf{r}) \nabla u^i(\mathbf{r})\cdot\nabla u^j(\mathbf{r}) \,\mathrm{d}\mathbf{r} \approx V_i^j[\sigma],
\end{equation}
where $i,j\in\{1,2,\ldots,E\}$ and $i\notin\{j-1,j,j+1\}$, with the electrode indices interpreted cyclically. Let
$\sigma(\mathbf{r}) = \sigma_0(\mathbf{r})+\delta\sigma(\mathbf{r})$, where $\sigma_0\in C^1(\Omega)$ is a known reference conductivity and $\delta\sigma$ denotes the conductivity perturbation. By linearizing \eqref{eq:reciprocity} about $\sigma_0$, we obtain
\begin{equation}\label{eq:reciprocity_delta}
    \frac{1}{I} \int_{\Omega} \delta\sigma(\mathbf{r}) \nabla u_0^i(\mathbf{r})\cdot\nabla u_0^j(\mathbf{r})
    \,\mathrm{d}\mathbf{r} \approx \delta V_i^j,
\end{equation}
where $u_0^j$ denotes the solution of \eqref{eq:forward} with $\sigma$ replaced by $\sigma_0$, and
$\delta V_i^j := V_i^j[\sigma]-V_i^j[\sigma_0]$.

By partitioning $\Omega$ into $N$ finite elements, $\overline{\Omega}=\bigcup_{k=1}^{N}\overline{T_k},$
and approximating the conductivity perturbation as piecewise constant,
$\delta\sigma(\mathbf{r}) \approx \sum_{k=1}^{N}\delta\sigma_k\chi_{T_k}(\mathbf{r}),$
the $E(E-3)$ linearized equations in
\eqref{eq:reciprocity_delta} can be expressed in matrix form as
\begin{equation}\label{eq:linearized_system}
    \mathbf{S}\boldsymbol{\delta\sigma} = \boldsymbol{\delta V},
\end{equation}
where
$\boldsymbol{\delta\sigma}
    :=
    \begin{pmatrix}
        \delta\sigma_1 &
        \delta\sigma_2 &
        \cdots &
        \delta\sigma_N
    \end{pmatrix}^{T}
    \in\mathbb{R}^{N}
$
is the discretized conductivity perturbation. Let
$\mathcal{M}
    :=
    \left\{
        (i,j)\in\{1,2,\ldots,E\}^{2}
        :
        i\notin\{j-1,j,j+1\}
    \right\},
$
where the electrode indices are interpreted cyclically. The set
$\mathcal{M}$ contains all admissible current-injection and
voltage-measurement pairs and satisfies
$|\mathcal{M}|=E(E-3)$. After arranging the elements of $\mathcal{M}$
in a fixed order, the data vector is defined by
\begin{equation}\label{eq:data_vector}
    \boldsymbol{\delta V}
    :=
    \left(
        \delta V_i^j
    \right)_{(i,j)\in\mathcal{M}}
    \in\mathbb{R}^{E(E-3)}.
\end{equation}
The sensitivity matrix
$\mathbf{S}\in\mathbb{R}^{E(E-3)\times N}$ is given by
\begin{equation}\label{eq:sensitivity_matrix}
    \mathbf{S}
    =
    \begin{pmatrix}
        \mathbf{S}_1 &
        \mathbf{S}_2 &
        \cdots &
        \mathbf{S}_N
    \end{pmatrix},
\end{equation}
where the $k$-th column is
$\mathbf{S}_k
    :=
    \left(
        s_k^{i,j}
    \right)_{(i,j)\in\mathcal{M}},
    \qquad k=1,2,\ldots,N,$
with
\begin{equation}\label{eq:sensitivity_entry}
    s_k^{i,j}
    :=
    -\frac{1}{I}
    \int_{T_k}
    \nabla u_0^i(\mathbf{r})
    \cdot
    \nabla u_0^j(\mathbf{r})
    \,\mathrm{d}\mathbf{r}.
\end{equation}
Therefore, the inverse problem of EIT consists of recovering the
conductivity perturbation vector $\boldsymbol{\delta\sigma}$, and
hence the conductivity distribution
$\boldsymbol{\sigma}
=
\boldsymbol{\sigma}_0+\boldsymbol{\delta\sigma}$,
from the known domain geometry, electrode configuration,
current-injection patterns, and measured voltage-difference data
$\boldsymbol{\delta V}$.

To mitigate the ill-posedness of the inverse problem, the conductivity
perturbation $\boldsymbol{\delta\sigma}$ is approximated by a minimizer
of the following regularized data-fitting problem:
\begin{equation}\label{problem:minimizing_general}
    \boldsymbol{\delta\sigma}^{\lambda}
    =
    \underset{\boldsymbol{\delta\sigma}\in\mathbb{R}^{N}}
    {\operatorname{arg\,min}}
    \left\{
        \left\|
            \mathbf{S}\boldsymbol{\delta\sigma}
            -
            \boldsymbol{\delta V}
        \right\|_{\ell_2}^{2}
        +
        \lambda\operatorname{Reg}
        \left(\boldsymbol{\delta\sigma}\right)
    \right\}.
\end{equation}
Here, the first and second terms in the objective functional represent
the data-fidelity and regularization terms, respectively, while
$\lambda>0$ is the regularization parameter that controls the trade-off
between data fidelity and regularization.

\subsection{Self-Generated Local Structural Prior}

We first introduce the Gaussian similarity matrix $\mathbf{C}$ proposed in \cite{Lv2020}. Specifically, the matrix
$\mathbf{C}=(c_{ij})_{i,j=1}^{N}\in\mathbb{R}^{N\times N}$ is defined by \begin{equation}\label{eq:Gaussian_similarity_matrix}
    c_{ij}
    =
    \exp\left(
        -\frac{
            \left|
                \boldsymbol{\delta\sigma}^{\lambda}[i]
                -
                \boldsymbol{\delta\sigma}^{\lambda}[j]
            \right|^{2}
        }{h^{2}}
    \right),
    \qquad i,j=1,2,\ldots,N,
\end{equation}
where $\boldsymbol{\delta\sigma}^{\lambda}$ is the preliminary reconstruction obtained by solving
\eqref{problem:minimizing_general}, and $h>0$ is the kernel-width parameter determined as follows. First, a prescribed region $D\Subset\Omega$ located away from the boundary $\partial\Omega$ is selected. The standard deviation of the reconstructed conductivity perturbation within $D$, denoted by $\operatorname{std}(\boldsymbol{\delta\sigma}^{\lambda}|_{D})$, is then computed. The kernel width is defined as
\begin{equation}\label{eq:kernel_width}
    h = \alpha\,
    \operatorname{std}
    \left(
        \boldsymbol{\delta\sigma}^{\lambda}|_{D}
    \right),
\end{equation}
where $\alpha>0$ is a constant independent of the object being imaged and the experimental setting, but dependent on the regularization parameter $\lambda$ and the mesh size.

Next, we compute the eigendecomposition of the Gaussian similarity matrix $\mathbf{C}$:
\begin{equation}\label{eq:eigendecomposition_C}
    \mathbf{C} = \mathbf{P}\boldsymbol{\Lambda}\mathbf{P}^{T},
\end{equation}
where $\boldsymbol{\Lambda} = \operatorname{diag}(\nu_1,\nu_2,\ldots,\nu_N)$
is the diagonal matrix of eigenvalues of $\mathbf{C}$, ordered such that $\nu_1\geq\nu_2\geq\cdots\geq\nu_N$, and $\mathbf{P} =
    \begin{pmatrix}
        \mathbf{v}_1 & \mathbf{v}_2 & \cdots & \mathbf{v}_N
    \end{pmatrix}$
is the corresponding orthogonal eigenvector matrix. Here, $\mathbf{v}_k$ denotes the eigenvector associated with the eigenvalue $\nu_k$.

We then select the eigenvector whose standard deviation over the prescribed region $D$ is minimal. Specifically, let
\begin{equation}\label{eq:selected_eigenvector}
    k^* =
    \underset{k\in\{1,2,\ldots,K\}}{\argmin}\,
    \operatorname{std}\left(\mathbf{v}_k|_D\right).
\end{equation}
Here, $K\in \mathbb{N}_+$ is the number of leading eigenvectors considered.
The selected eigenvector $\mathbf{v}_{k^*}$ encodes structural information within $\Omega$ and provides better foreground-background separability than the preliminary reconstruction $\boldsymbol{\delta\sigma}^{\lambda}$.

A segmentation procedure is then applied to $\mathbf{v}_{k^*}$ to separate $\Omega$ into foreground and background regions. The resulting binary mask $\mathbf{M}$ is defined by
\begin{equation}\label{eq:segmentation_mask}
    \mathbf{M}(\mathbf{r}) =
    \begin{cases}
        1, & \mathbf{r}\in\mathcal{F},\\
        0, & \mathbf{r}\in\mathcal{B},
    \end{cases}
    \qquad
    \mathcal{F}\cup\mathcal{B}=\Omega,
    \quad
    \mathcal{F}\cap\mathcal{B}=\varnothing,
\end{equation}
where $\mathcal{F}$ and $\mathcal{B}$ denote the foreground and background regions, respectively. The foreground generally consists of the organs or structures of primary interest. For example, in human lung EIT, the foreground typically comprises the two lungs and the heart, whereas the background consists of the remaining regions within the imaging domain.

Using the segmentation mask, we define $\widetilde{\boldsymbol{\delta\sigma}}$ as a mean-prior conductivity perturbation. Specifically, let
\begin{equation}\label{eq:foreground_mean}
    \overline{\boldsymbol{\delta\sigma}}_{\mathcal{F}}
    :=
    \operatorname{mean}
    \left(
        \boldsymbol{\delta\sigma}^{\lambda}\big|_{\mathcal{F}}
    \right)
\end{equation}
denote the regional mean of the preliminary reconstruction over the foreground region. The mean-prior conductivity perturbation is then defined by
\begin{equation}\label{eq:mean_prior}
    \widetilde{\boldsymbol{\delta\sigma}}(\mathbf{r})
    :=
    \begin{cases}
        \overline{\delta\sigma}_{\mathcal{F}},
        & \mathbf{r}\in \mathcal{F},\\
        0,
        & \mathbf{r} \in \mathcal{B}.
    \end{cases}
\end{equation}
Equivalently, the mean-prior conductivity perturbation can be written as $\widetilde{\delta\sigma}(\mathbf{r}) =  M(\mathbf{r})\,\overline{\delta\sigma}_{\mathcal{F}},$ where $\overline{\delta\sigma}_{\mathcal{F}}$ denotes the mean conductivity perturbation over the foreground region. In the discrete setting, this expression is equivalently written as $\widetilde{\boldsymbol{\delta\sigma}} =  \mathbf{M}\odot \overline{\boldsymbol{\delta\sigma}}_{\mathcal{F}},$
where $\odot$ denotes the Hadamard (componentwise) product.

Using the mask $\mathbf{M}$ defined above, we introduce the following regularization functional, referred to as the self-generated local structural prior (SGLSP):
\begin{equation}\label{eq:SGLSP}
\begin{aligned}
    \operatorname{Reg}_{\mathrm{SGLSP}}
    \left(\boldsymbol{\delta\sigma}\right)
    ={}&
    \frac{\lambda_1}{2\lambda}
    \left\|
        \mathbf{D}
        \left(
            \mathbf{M}\odot\boldsymbol{\delta\sigma}
        \right)
    \right\|_{\ell_2}^{2}
    \\[2mm]
    &+
    \frac{\lambda_2}{2\lambda}
    \left\|
        \mathbf{D}
        \left[
            \mathbf{M}\odot
            \left(
                \boldsymbol{\delta\sigma}
                -
                \widetilde{\boldsymbol{\delta\sigma}}
            \right)
        \right]
    \right\|_{\ell_2}^{2}.
\end{aligned}
\end{equation}
The first term promotes spatial smoothness within the region identified by the mask, whereas the second term encourages the reconstruction to conform locally to the `mean prior'
$\widetilde{\boldsymbol{\delta\sigma}}$.

By combining the local structural prior defined above with a global
regularization term, such as the nonlinear weighted anisotropic total
variation (NWATV) regularization proposed in \cite{Song2022}, we
approximate $\boldsymbol{\delta\sigma}$ by a minimizer of the following
regularized data-fitting problem:
\begin{equation}\label{minimization:final_regularized}
\begin{split}
 &\delta{\bm \sigma}^{\lambda,\lambda_1,\lambda_2} = \\
 & \quad  \argmin_{\delta{\bm \sigma}}\Big\{\frac{1}{2}\|\mathbf{S}\delta{\bm \sigma} - \delta\mathbf{V}\|_{\ell_2}^2 +\frac{\lambda_1}{2} \|\mathbf{D}(\mathbf{M} \odot \delta{\bm \sigma})\|_{\ell_2}^2 \\
  & \quad  + \frac{\lambda_2}{2} \|\mathbf{D}(\mathbf{M}\odot(\delta{\bm \sigma}-\delta\tilde{\bm \sigma}))\|_{\ell_2}^2 +\lambda \|\mathbf{p}\odot \mathbf{D}\delta{\bm \sigma}\|_{\ell_1} \Big\}.
\end{split}
\end{equation}
Here, $\mathbf{D} = [\mathbf{D}_x; \mathbf{D}_y] \in \mathbb{R}^{2N \times N}$ denotes the discrete gradient operator, where $\mathbf{D}_x, \mathbf{D}_y \in \mathbb{R}^{N \times N}$ are respectively the first-order difference matrices along the $x$- and $y$-directions. The vector $\mathbf{p} = [\bm{\zeta}(\delta{\bm \sigma}); \bm{\zeta}(\delta{\bm \sigma})] \in \mathbb{R}^{2N}$ is the nonlinear weights, where $\bm{\zeta}\in \mathbb{R}^{N}$ is defined componentwise by 
${\bm{\zeta}}(\delta{\bm \sigma})[k] = 1/(|\mathbf{D}_x\delta{\bm \sigma}[k]|^2 + |\mathbf{D}_y\delta{\bm \sigma}[k]|^2 + \beta)$,
where $\beta$ is a small positive parameter introduced to avoid division by zero.

\subsection{Reconstruction Algorithm}

To solve the minimization problem
\eqref{minimization:final_regularized}, we propose an ADMM-type reconstruction algorithm \cite{Boyd2011}. Introducing the auxiliary variable $\mathbf{z}=\mathbf{D}\boldsymbol{\delta\sigma},$
we define the augmented Lagrangian
\begin{equation}
\begin{split}
  & \mathcal{L}_\rho(\delta{\bm \sigma},\mathbf{z},\mathbf{p};\mathbf{y}) = \frac{1}{2}\|\mathbf{S}\delta{\bm \sigma} - \delta\mathbf{V}\|_{\ell_2}^2 +\frac{\lambda_1}{2} \|\mathbf{D}(\mathbf{M} \odot \delta{\bm \sigma})\|_{\ell_2}^2 \\
  &\qquad\qquad  + \frac{\lambda_2}{2} \|\mathbf{D}(\mathbf{M}\odot(\delta{\bm \sigma}-\widetilde{\delta{\bm \sigma}}))\|_{\ell_2}^2 +\lambda \|\mathbf{p}\odot \mathbf{z}\|_{\ell_1} \\
  & \qquad\qquad +\frac{\rho}{2}\|\mathbf{D}\delta{\bm \sigma}-\mathbf{z}\|^2_{\ell_2} + \mathbf{y}^T(\mathbf{D}\delta{\bm \sigma}-\mathbf{z}).
\end{split}
\end{equation}
where $\mathbf{y}$ is the Lagrange multiplier associated with the constraint $\mathbf{D}\boldsymbol{\delta\sigma}=\mathbf{z}$, and $\rho>0$ is the penalty parameter.

Starting from an initial iterate $(\mathbf{z}_0,\mathbf{p}_0,\mathbf{y}_0)$, the reconstruction $\boldsymbol{\delta\sigma}^{\lambda,\lambda_1,\lambda_2}$ is computed iteratively according to
\begin{subequations}\label{eq:ADMM_updates}
\begin{align}
    \delta{\bm \sigma}_{n+1} &= \argmin_{\delta{\bm \sigma}} \mathcal{L}_\rho(\delta{\bm \sigma}, \mathbf{z}_n, \mathbf{p}_n; \mathbf{y}_n) \label{eq:update_8a} \\
    \mathbf{z}_{n+1} &= \argmin_{\mathbf{z}} \mathcal{L}_\rho(\delta{\bm \sigma}_{n+1}, \mathbf{z}, \mathbf{p}_n; \mathbf{y}_n) \label{eq:update_8b} \\
    \mathbf{p}_{n+1} &= [\bm{\zeta}(\delta{\bm \sigma}_{n+1}); \bm{\zeta}(\delta{\bm \sigma}_{n+1})] \label{eq:update_8c} \\
    \mathbf{y}_{n+1} &= \mathbf{y}_n + \rho(\mathbf{D}\delta{\bm \sigma}_{n+1} - \mathbf{z}_{n+1}). \label{eq:update_8d}
\end{align}
\end{subequations}
Here, the update of $\mathbf{p}$ corresponds to an iterative reweighting step, since the nonlinear weights depend on the current estimate of the conductivity perturbation. The iterations are terminated when a prescribed stopping criterion is satisfied, and the final iterate is taken as the reconstructed conductivity perturbation $\boldsymbol{\delta\sigma}^{\lambda,\lambda_1,\lambda_2}$.

Note that the minimizers \eqref{eq:update_8a} and \eqref{eq:update_8b} have the following closed form
\begin{equation}
    \begin{aligned}
    & \delta{\bm \sigma}_{n+1} = \left( \frac{1}{\rho}\mathbf{S}^T \mathbf{S} + \mathbf{D}^T \mathbf{D} + \frac{\lambda_1 + \lambda_2}{\rho}\mathbf{M}\mathbf{D}^T\mathbf{D}\mathbf{M} \right)^{-1} \cdot \\
    &\quad \left( \frac{1}{\rho}\mathbf{S}^T \delta\mathbf{V} + \frac{\lambda_2}{\rho}\mathbf{M}\mathbf{D}^T\mathbf{D}\mathbf{M}\delta\tilde{\bm \sigma} + \mathbf{D}^T \mathbf{z}_n - \mathbf{D}^T \frac{\mathbf{y}_n}{\rho} \right),
    \end{aligned}
    \label{eq:solve_10}
\end{equation}
and
\begin{equation}
    \mathbf{z}_{n+1}[k] = h_{\frac{\lambda \mathbf{p}_n[k]}{\rho}} \left(\mathbf{D}\delta{\bm \sigma}_{n+1}[k] + \frac{\mathbf{y}_n[k]}{\rho}\right).
    \label{eq:solve_11}
\end{equation}
Here, $\mathbf{y}_n[k]$ and $\mathbf{z}_{n+1}[k]$ are the $k$-th elements of $\mathbf{y}_n$ and $\mathbf{z}_{n+1}$, respectively, for $k = 1, 2, \dots, 2N$. $h_g(\cdot)$ is the soft thresholding operator, defined as:
\begin{equation}
    h_g(f) = 
    \begin{cases} 
    0, & |f| \le g \\
    f - g\,\mathrm{sgn}(f), & \text{otherwise} 
    \end{cases}
    \label{eq:solve_12}
\end{equation}
where $g\geq 0$ and $\operatorname{sgn}(\cdot)$ is the sign function.

We conclude this section by presenting pseudocode that summarizes the complete reconstruction procedure described above.

\begin{algorithm}[H]
\caption{SGLSP-ADMM Reconstruction Algorithm}
\begin{algorithmic}[1]
\renewcommand{\algorithmicrequire}{\textbf{Input:}}
\renewcommand{\algorithmicensure}{\textbf{Output:}}
\REQUIRE Measurement data $\boldsymbol{\delta V}$, sensitivity matrix $\mathbf{S}$, mask matrix $\mathbf{M}$, mean-prior conductivity perturbation $\widetilde{\boldsymbol{\delta\sigma}}$, parameters $\lambda, \rho, \lambda_1, \lambda_2$, $\beta$, max iterations $K_{\max}$, tolerance $tol$.
\ENSURE Reconstructed conductivity perturbation $\boldsymbol{\delta\sigma}$
\STATE \textbf{Initialize:} $\boldsymbol{\delta\sigma}_0 = \mathbf{0}$, $\mathbf{y}_0 = \mathbf{0}$, $\mathbf{z}_0 = \mathbf{0}$, $\mathbf{p}_0 = \mathbf{1}$, $n = 1$
\WHILE{$n \le K_{\max}$}
    \STATE Update $\boldsymbol{\delta\sigma}_{n+1}$ by solving the consolidated linear system equation \eqref{eq:solve_10}.
    \STATE Update auxiliary variable $\mathbf{z}_{n+1}$ using the soft thresholding formula \eqref{eq:solve_11}.
    \STATE Update the nonlinear local gradient weights $\mathbf{p}_{n+1}$ using formula \eqref{eq:update_8c}.
    \STATE Update the global Lagrange multiplier $\mathbf{y}_{n+1}$ using formula \eqref{eq:update_8d}.

    \IF{$\|\boldsymbol{\delta\sigma}_{n+1} - \boldsymbol{\delta\sigma}_n\|_{l_2} \le tol$}
        \STATE \textbf{break}
    \ENDIF
    \STATE $n \leftarrow n + 1$
\ENDWHILE
\end{algorithmic}
\end{algorithm}

\section{Results}

To evaluate the performance of the proposed algorithm, we conducted numerical and water tank imaging experiments.

\subsection{General Experimental Setup}

In the experiments, we placed 16 electrodes as uniformly as possible on the boundary of the tested object. We used the adjacent-stimulation-adjacent-measurement protocol in all experiments: we injected a driving current with an amplitude of $1\,\mathrm{mA}$ and measured the boundary voltages between adjacent electrode pairs.

We compared the proposed algorithm with the Tikhonov, Isotropic TV implemented by PDIPM \cite{Zhou2015}, Anisotropic TV implemented by ADMM \cite{Boyd2011}, and NWATV. For quantitative evaluation, we computed relative error (RE), peak signal-to-noise ratio (PSNR), and structural similarity index (SSIM) in the numerical experiments to evaluate the accuracy of reconstructed images. At the $n$-th iteration, $\mathrm{RE}(n)$ is defined as
\begin{equation}
    \mathrm{RE}(n)=\frac{\|\bm{\sigma}^{n}-\bm{\sigma}^{*}\|_2}{\|\bm{\sigma}^{*}\|_2},
\end{equation}
and $\mathrm{PSNR}(n)$ is defined as \cite{Liu2023PET}
\begin{equation}
    \mathrm{PSNR}(n)=10\log_{10}\left(\frac{\max(\bm{\sigma}^{*}\odot\bm{\sigma}^{*})}{\mathrm{MSE}(n)}\right).
\end{equation}
Here, $\odot$ denotes elementwise vector multiplication, $\bm{\sigma}^{*}$ denotes the true or reference conductivity distribution, $\bm{\sigma}^{n}$ denotes the reconstructed conductivity obtained at the $n$-th iteration, and $\mathrm{MSE}(n)$ is the mean squared error between $\bm{\sigma}^{*}$ and $\bm{\sigma}^{n}$:
\begin{equation}
    \mathrm{MSE}(n)=\frac{1}{N}\sum_{i=1}^{N}\left[\bm{\sigma}^{n}(i)-\bm{\sigma}^{*}(i)\right]^2.
\end{equation}
The $\mathrm{SSIM}$ metric \cite{Wang2004} is further introduced to evaluate structural similarity and is defined as
\begin{equation*}
\begin{split}
    & \mathrm{SSIM}(\bm{\sigma}^{n},\bm{\sigma}^{*})=\\
    &
    \frac{[2 \mathbb{E}(\boldsymbol{\sigma}_n) \mathbb{E}(\boldsymbol{\sigma}_*)+C_1][2 \operatorname{Cov}(\boldsymbol{\sigma}_n,\boldsymbol{\sigma}_*)+C_2]}
    {\{[\mathbb{E}(\boldsymbol{\sigma}_n)]^2+[\mathbb{E}(\boldsymbol{\sigma}_*)]^2+C_1\}
    (\operatorname{Var}(\boldsymbol{\sigma}_n) + \operatorname{Var}(\boldsymbol{\sigma}_*)+C_2)}.
\end{split}
\end{equation*}
Here, $\mathbb{E}(\boldsymbol{\sigma}_n)$ and $\mathbb{E}(\boldsymbol{\sigma}_*)$ are the means of $\bm{\sigma}^{n}$ and $\bm{\sigma}^{*}$, respectively; $\operatorname{Var}(\boldsymbol{\sigma}_n)$ and $\operatorname{Var}(\boldsymbol{\sigma}_*)$ are their variances, respectively; and $\sigma_{n,*}$ is the covariance. In each experiment, we determined the stability constants by the dynamic range $L$ of the corresponding color scale, with $C_1=(0.01L)^2$ and $C_2=(0.03L)^2$. We computed RE, PSNR, and SSIM on a unified $256\times256$ interpolated image and valid-domain mask to reduce the influence of irregular finite-element mesh sizes on error evaluation. We determined statistics in $D$ according to the geometric centers of finite elements and included elements whose centers lay inside the boundary of $D$ in the statistics for subsequent computation of $\mathrm{std}(\boldsymbol{\delta\sigma}^\lambda|_D)$ and $\mathrm{std}(\mathbf{v}_{k^*}|_D)$.

We implemented all algorithms in MATLAB together with the EIDORS toolbox \cite{adler2006eidors}. We performed all computations and tests in this paper on a portable workstation (Lenovo Legion R7000, 2020 model) equipped with an AMD Ryzen 7 4800H processor and an RTX 1650 dedicated graphics card.

\subsection{Three-Dimensional Numerical Experiment}

\begin{figure}[htbp]
    \centering
    \begin{minipage}{0.28\columnwidth}
        \centering
        \includegraphics[width=\linewidth]{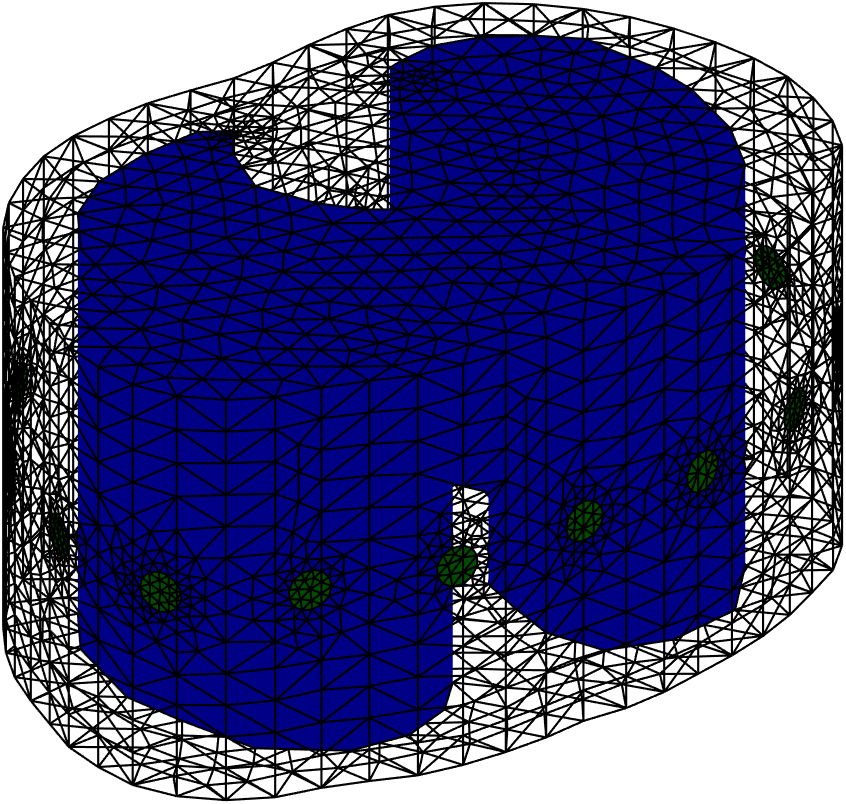}
        \vspace{-0.1cm}
        \centerline{\footnotesize (a)}
    \end{minipage}
    \hspace{0.05\columnwidth}
    \begin{minipage}{0.35\columnwidth}
        \centering
        \includegraphics[width=\linewidth]{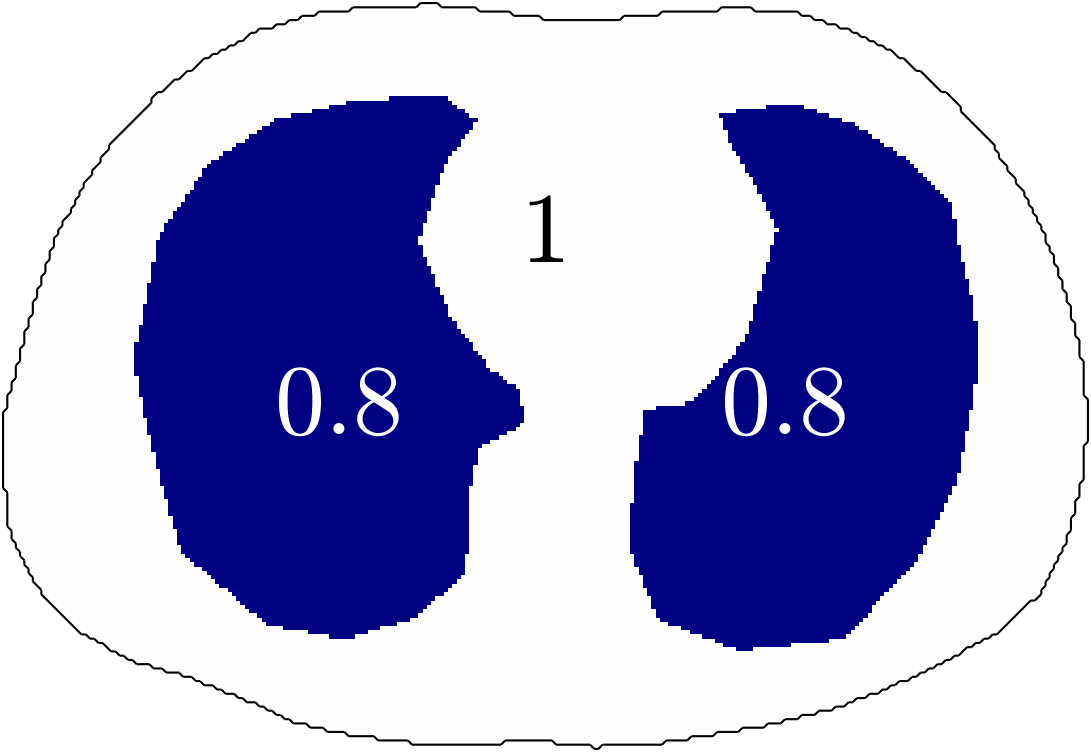}
        \vspace{-0.1cm}
        \centerline{\footnotesize (b)}
    \end{minipage}

    \vspace{0.1em}

    \begin{minipage}{0.33\columnwidth}
        \centering
        \includegraphics[width=\linewidth]{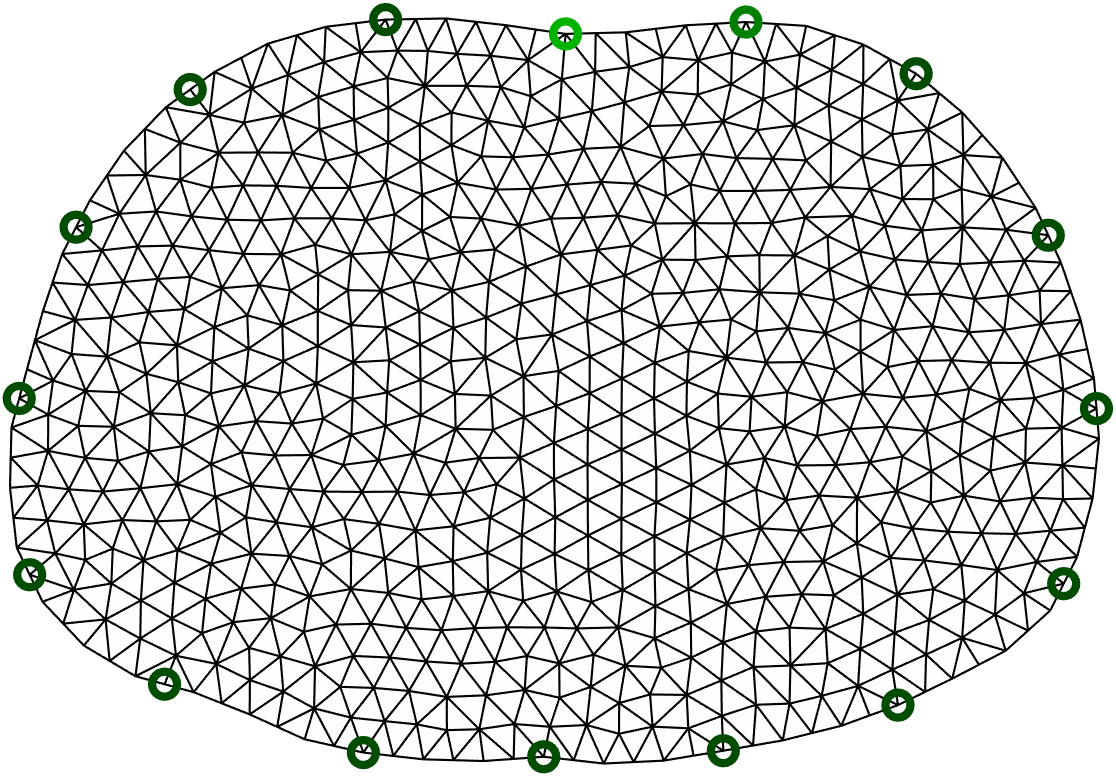}
        \vspace{-0.1cm}
        \centerline{\footnotesize (c)}
    \end{minipage}
    \hspace{0.05\columnwidth}
    \begin{minipage}{0.42\columnwidth}
        \centering
        \includegraphics[width=\linewidth]{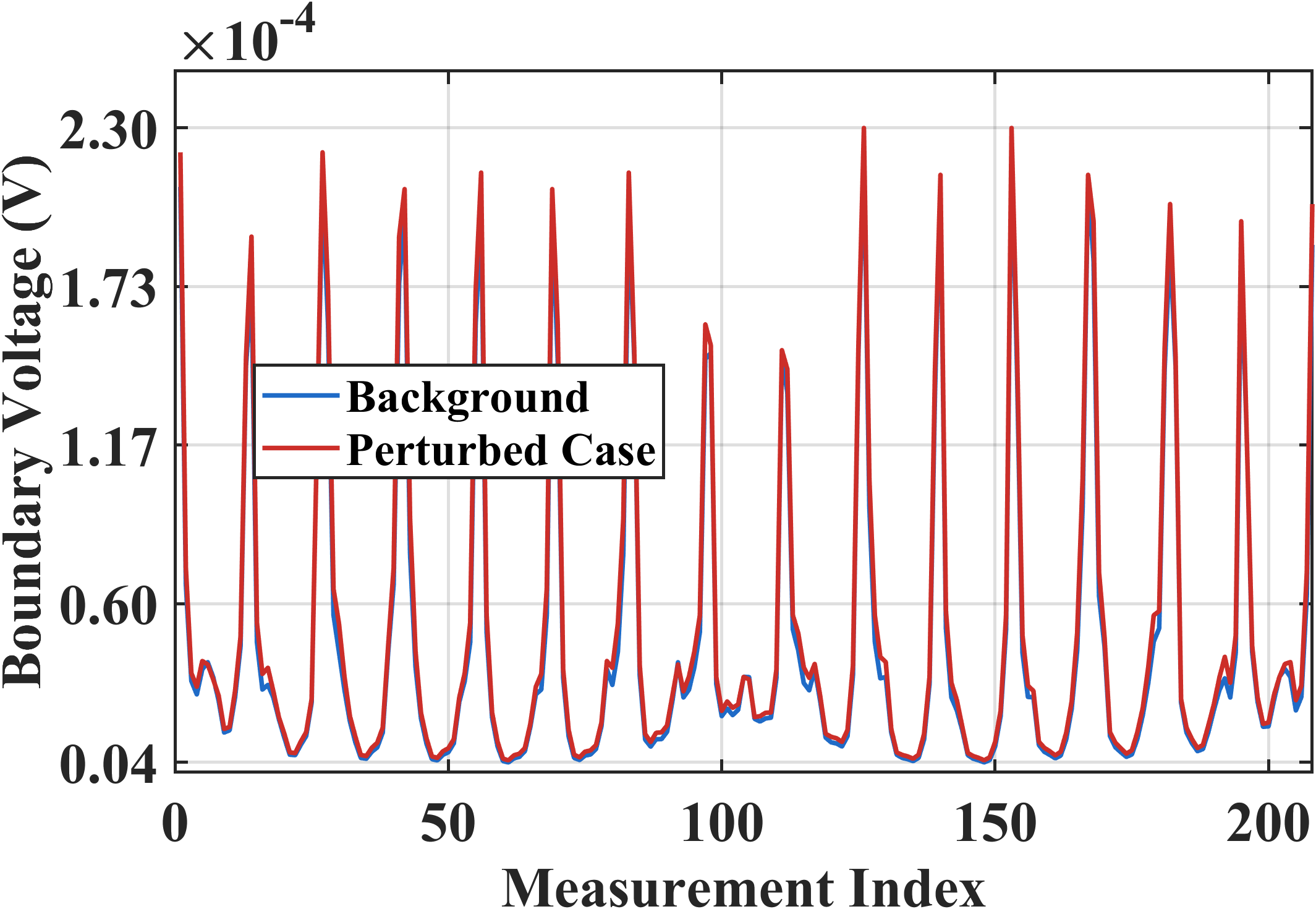}
        \vspace{-0.1cm}
        \centerline{\footnotesize (d)}
    \end{minipage}
    \caption{Three-dimensional numerical simulation model. (a) Three-dimensional forward mesh with electrodes. (b) True conductivity distribution on the central slice. (c) Triangular mesh used in the inverse reconstruction. (d) Boundary voltage vectors under the background (blue line) and target-perturbed states (red line); the inverse-problem input, $\delta \mathbf{V}$, is their difference.}
    \label{fig:3d_mesh_short}
\end{figure}

In this section, we used the three-dimensional adult thorax model in EIDORS for numerical simulation~\cite{adler2006eidors}. Fig. \ref{fig:3d_mesh_short}(a) shows the model geometry. We arranged 16 circular electrodes along the boundary of the central slice. The three-dimensional forward model contains 5379 nodes and 26358 tetrahedra. We set the background conductivity to $1.0$ S/m and the conductivity of the two lung anomaly regions to $0.8$ S/m. Fig. \ref{fig:3d_mesh_short}(b) shows the true conductivity distribution on the central slice, and Fig. \ref{fig:3d_mesh_short}(c) gives the two-dimensional central-slice mesh used for reconstruction, which contains 637 nodes and 1172 triangular elements.

We first solved the 3-D forward problem \eqref{eq:forward} to obtain the boundary voltage data, as shown in Fig.~\ref{fig:3d_mesh_short}(d). We then added Gaussian random noise at 50 dB and 30 dB to the clean simulated voltage-difference data to simulate EIT data under different measurement noise levels. The 50 dB noise level corresponds to a typical noise level of modern EIT systems~\cite{Bai2022Noise}, while the 30 dB case introduces additional noise to examine robustness under stronger noise. Based on the noisy simulated voltage-difference data, we used the proposed algorithm and four comparison methods to reconstruct the conductivity image of the central slice.

\begin{figure*}[!t]
    \centering
    \begin{minipage}{0.18\textwidth}
        \centering
        \includegraphics[width=\linewidth]{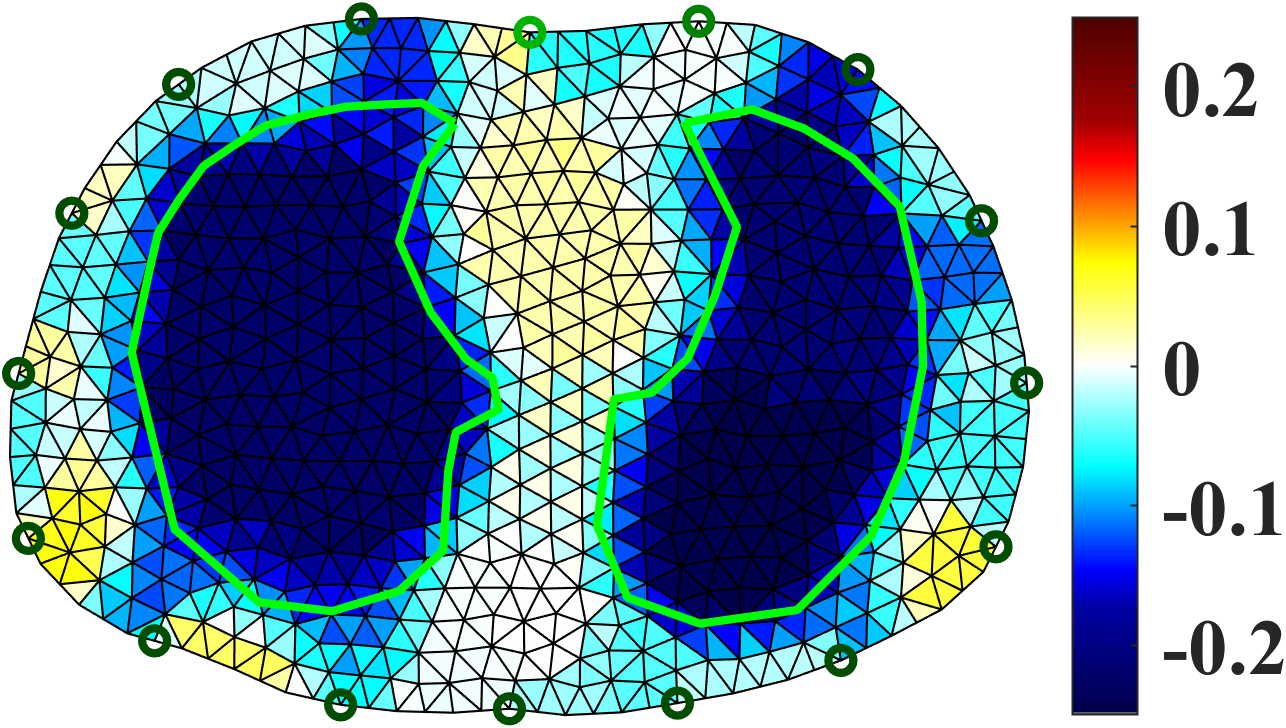}
        \centerline{\footnotesize (a)}
    \end{minipage}
    \hspace{0.01\textwidth}
    \begin{minipage}{0.18\textwidth}
        \centering
        \includegraphics[width=\linewidth]{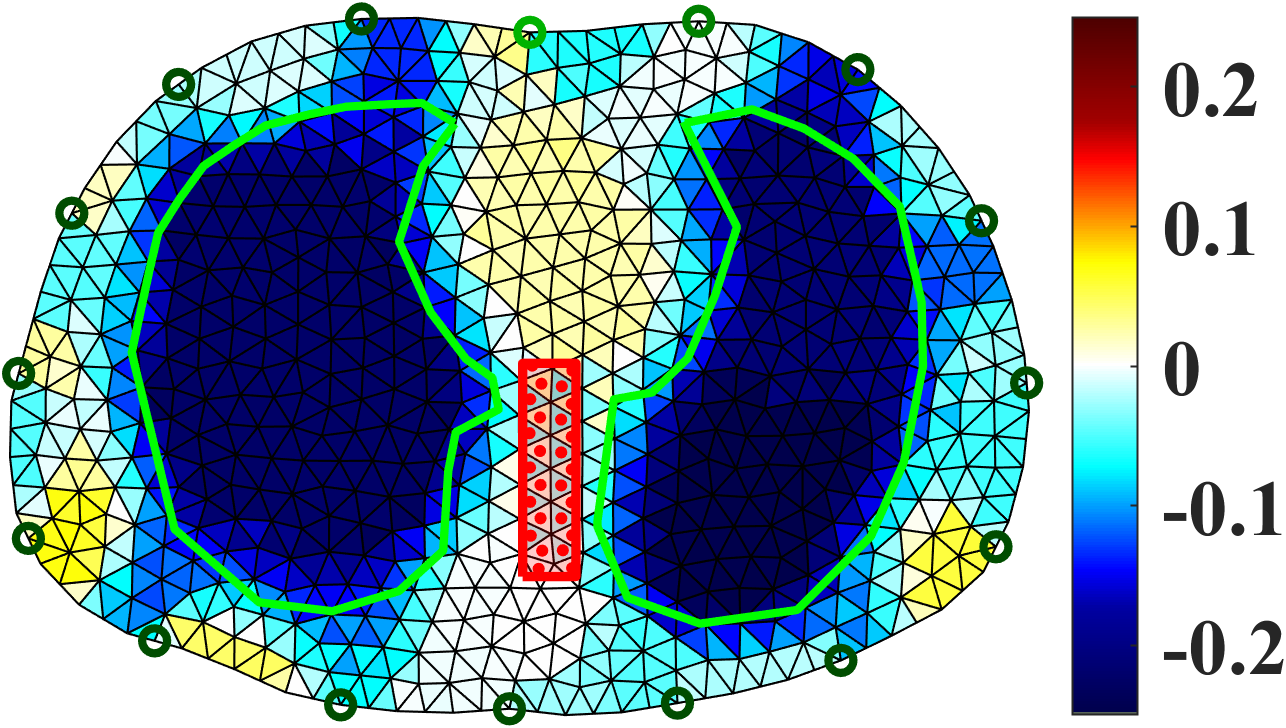}
        \centerline{\footnotesize (b)}
    \end{minipage}
    \hspace{0.01\textwidth}
    \begin{minipage}{0.18\textwidth}
        \centering
        \includegraphics[width=\linewidth]{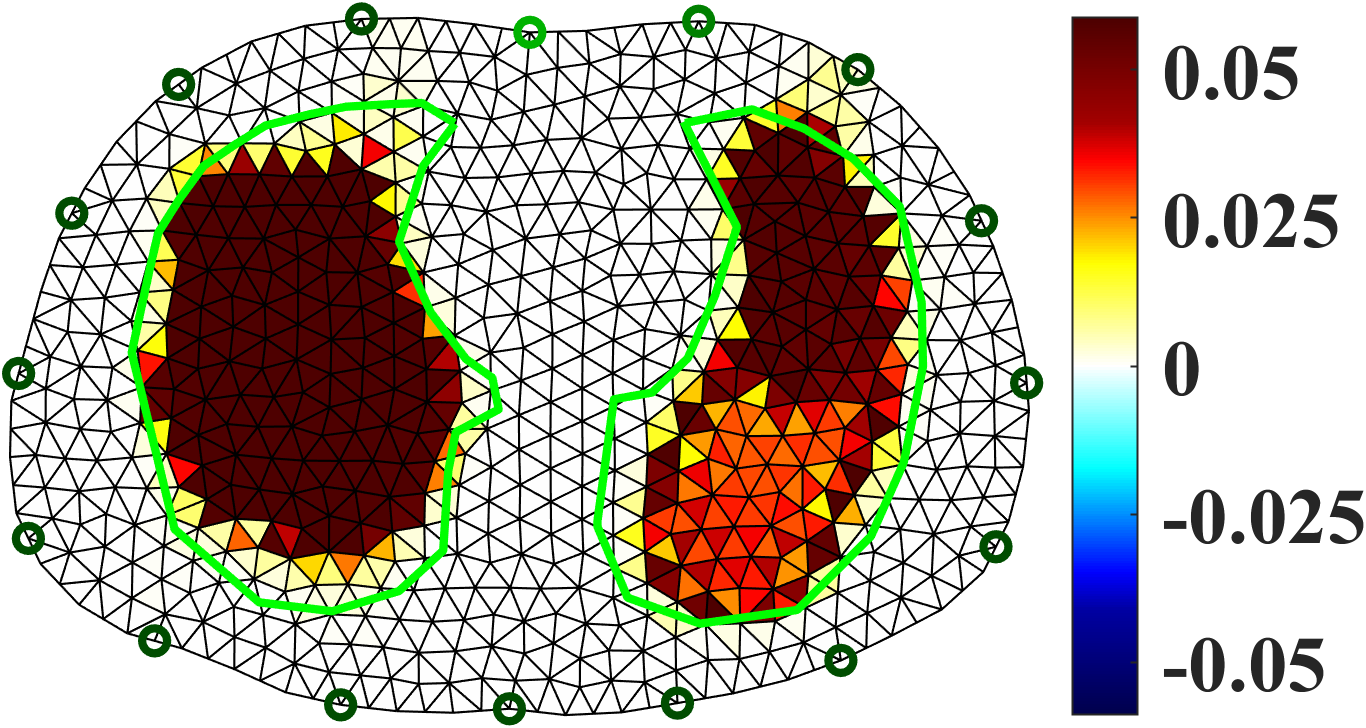}
        \centerline{\footnotesize (c)}
    \end{minipage}
    \hspace{0.01\textwidth}
    \begin{minipage}{0.145\textwidth}
        \centering
        \includegraphics[width=\linewidth]{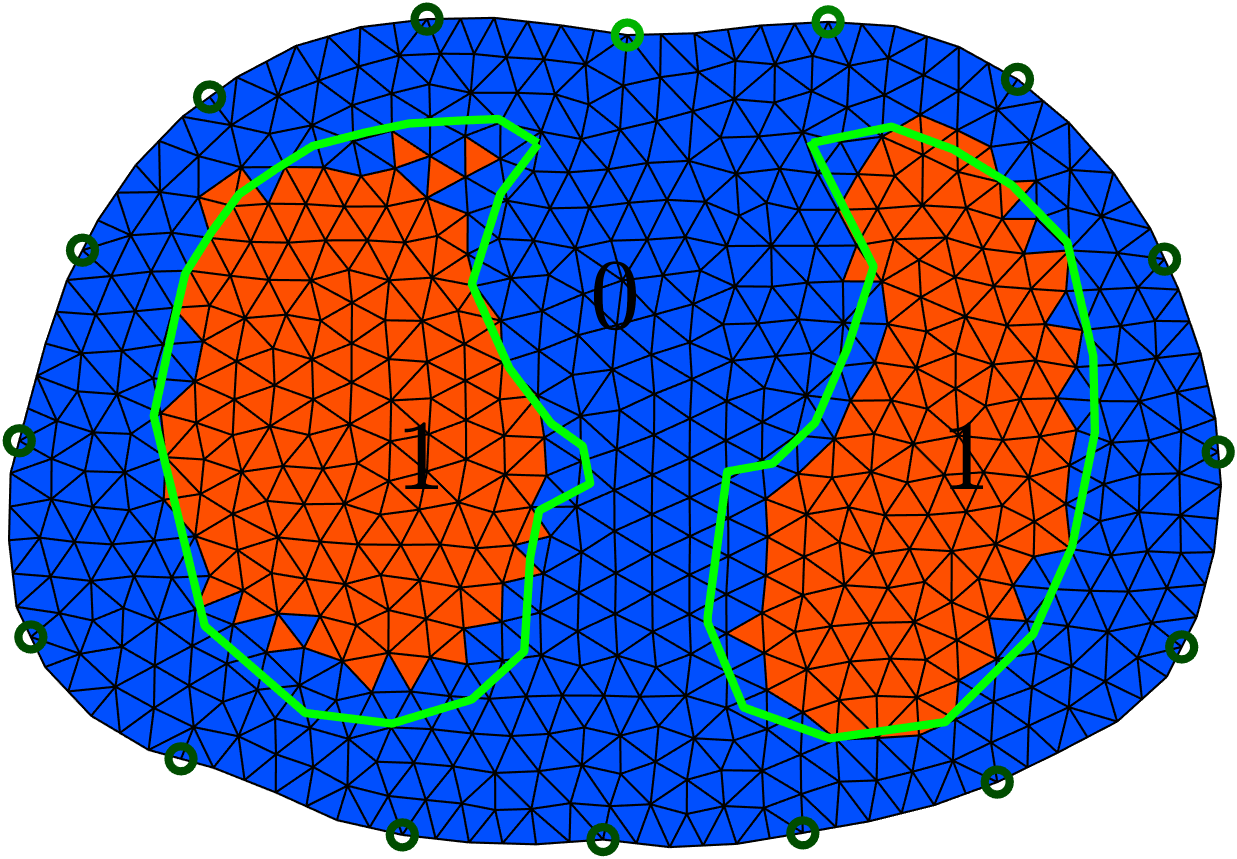}
        \centerline{\footnotesize (d)}
    \end{minipage}
    \hspace{0.01\textwidth}
    \begin{minipage}{0.18\textwidth}
        \centering
        \includegraphics[width=\linewidth]{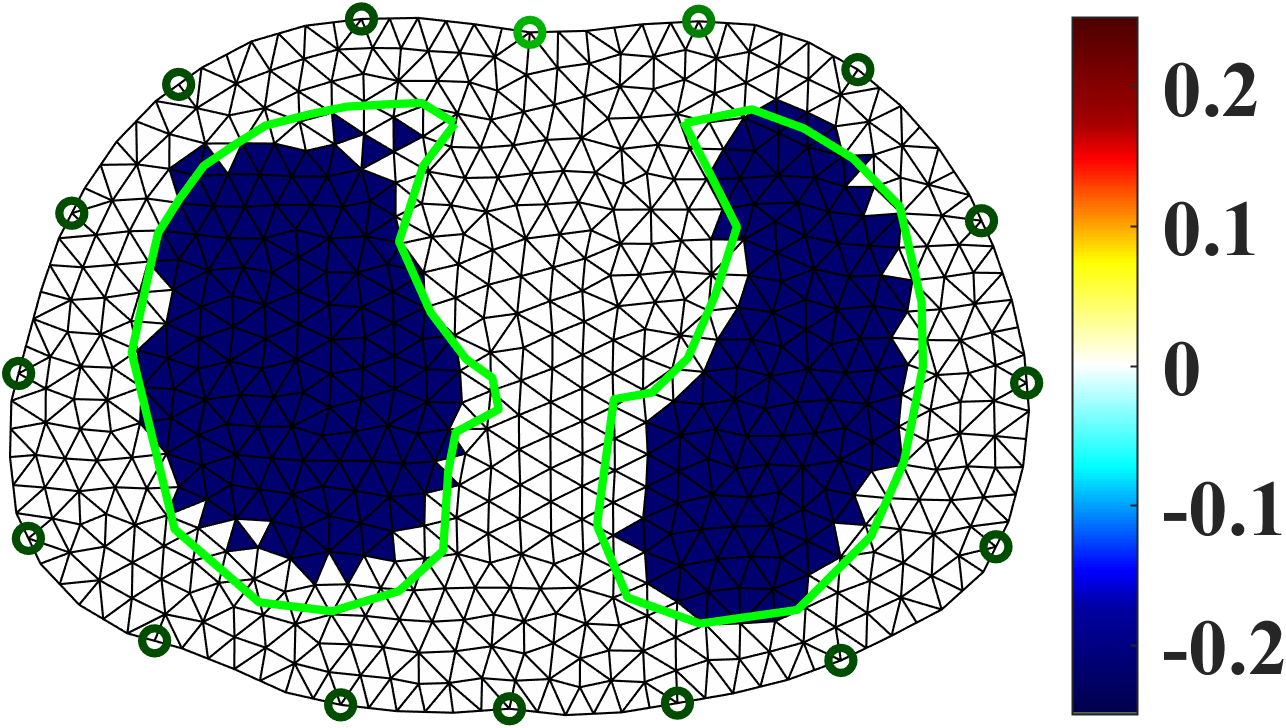}
        \centerline{\footnotesize (e)}
    \end{minipage}
    \caption{Extraction process of $\mathbf{M}$ and $\widetilde{\boldsymbol{\delta\sigma}}$ for the three-dimensional simulation, shown from (a) to (e). (a) $\boldsymbol{\delta\sigma}^\lambda$ obtained by using NWATV method. (b) Problematic region $D$ away from $\partial\Omega$. (c) Principal eigenvector of matrix $\mathbf{C}$ with the smallest standard deviation in $D$, namely the first principal eigenvector. (d) Mask $\mathbf{M}$ obtained by binary segmentation of the first principal eigenvector. (e) Assigned `mean prior' $\widetilde{\boldsymbol{\delta\sigma}}$.}
    \label{fig:3d_prior_roi_short}
\end{figure*}

\begin{figure}[H]
    \centering
    \begin{minipage}{0.35\columnwidth}
        \centering
        \includegraphics[width=\linewidth]{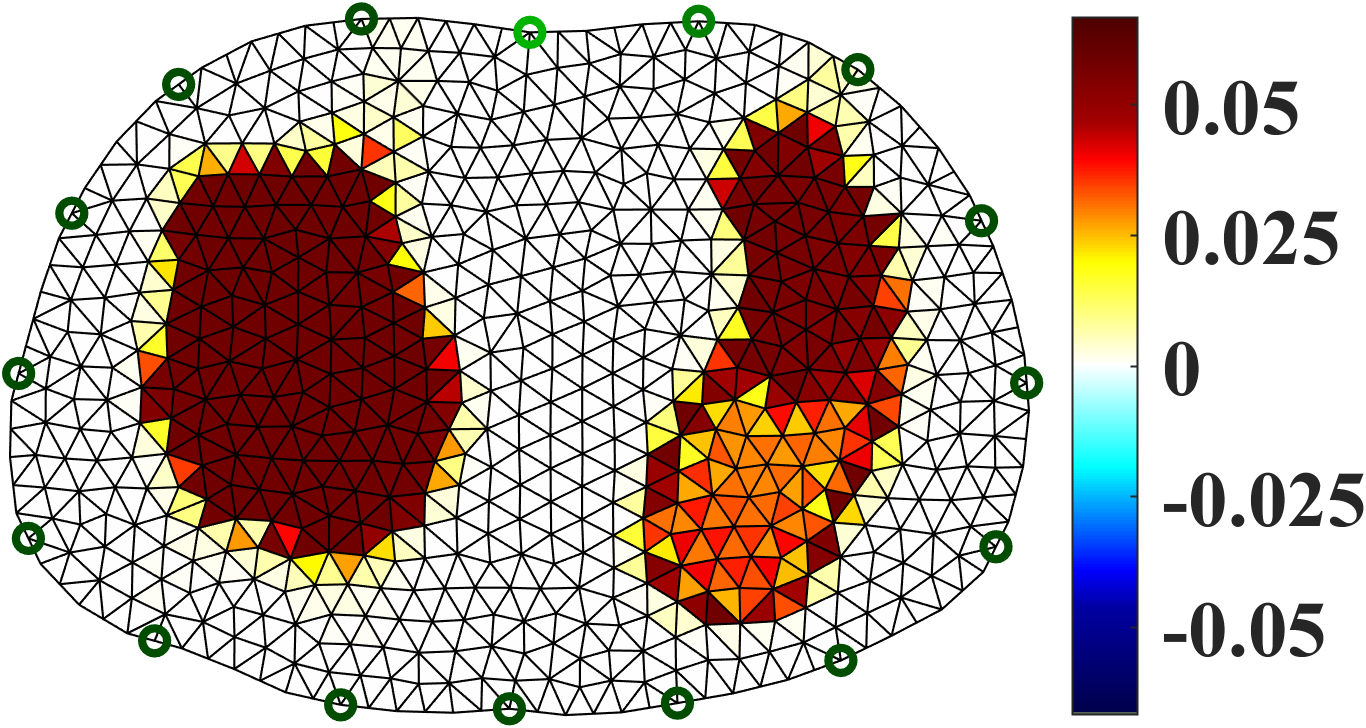}
        \centerline{\footnotesize (a)}
    \end{minipage}\hspace{0.06\columnwidth}
    \begin{minipage}{0.35\columnwidth}
        \centering
        \includegraphics[width=\linewidth]{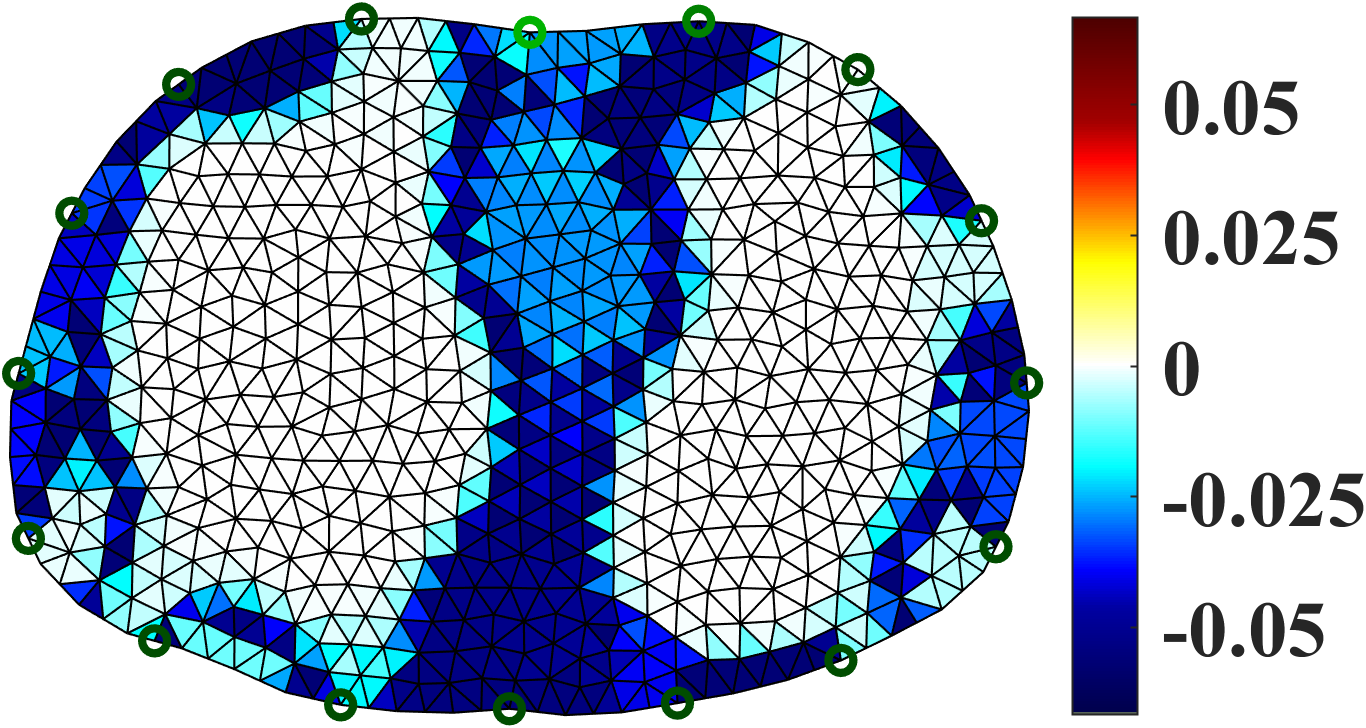}
        \centerline{\footnotesize (b)}
    \end{minipage}

    \vspace{0.1em}

    \begin{minipage}{0.35\columnwidth}
        \centering
        \includegraphics[width=\linewidth]{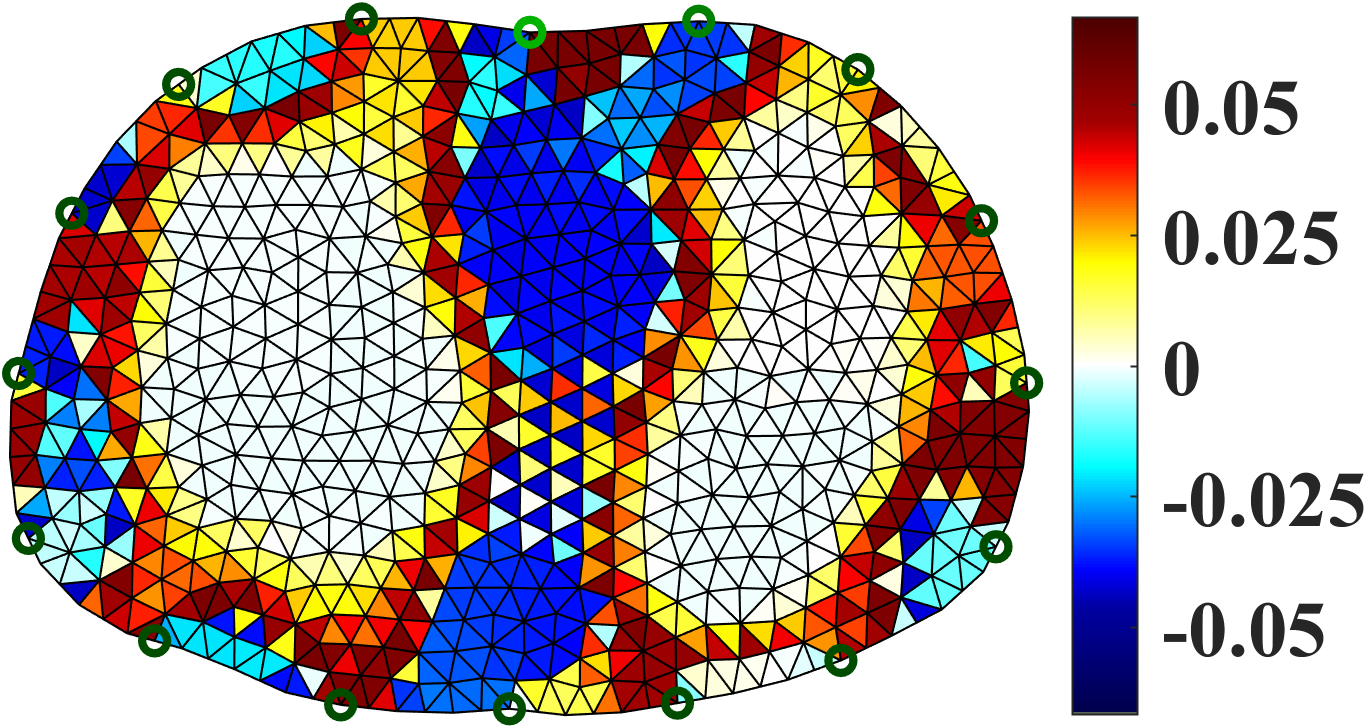}
        \centerline{\footnotesize (c)}
    \end{minipage}\hspace{0.06\columnwidth}
    \begin{minipage}{0.35\columnwidth}
        \centering
        \includegraphics[width=\linewidth]{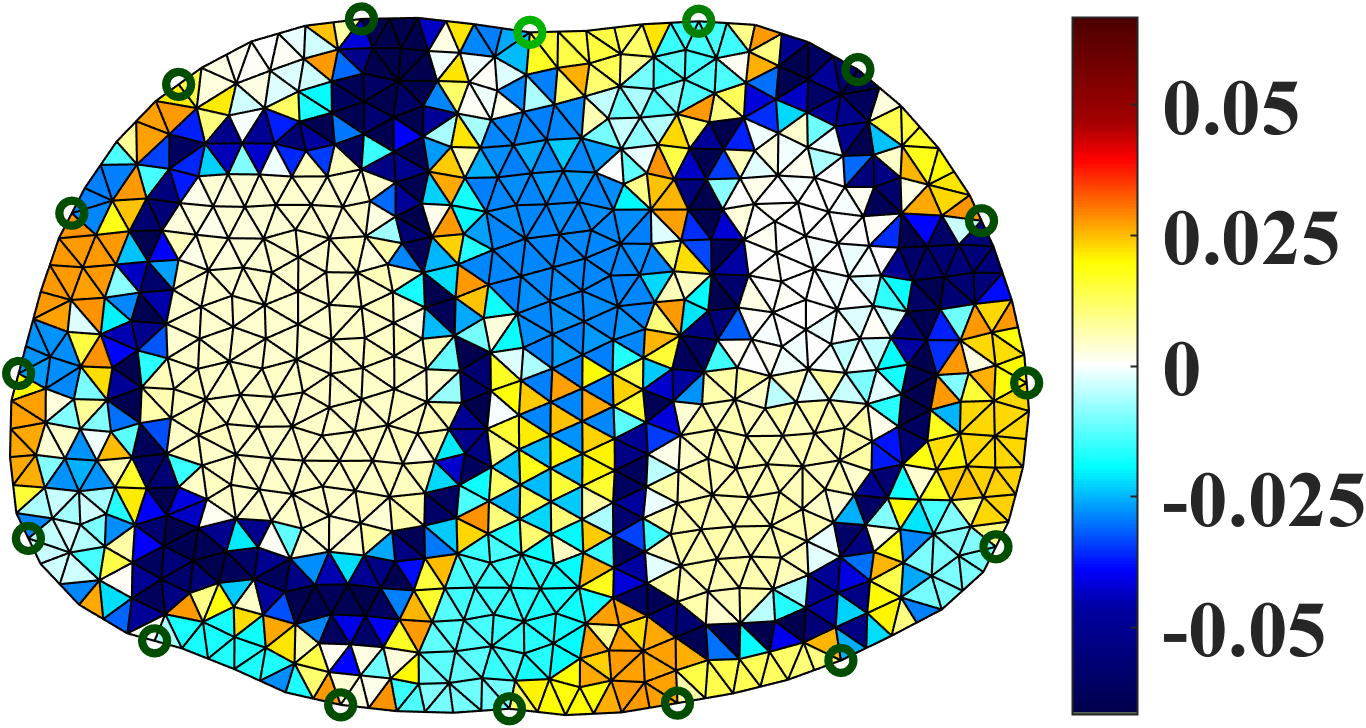}
        \centerline{\footnotesize (d)}
    \end{minipage}
    \caption{The first four principal eigenvectors of the Gaussian similarity matrix $\mathbf{C}$ constructed from the three-dimensional simulation NWATV reconstruction. (a)-(d) show the first to fourth principal eigenvectors, respectively, with corresponding eigenvalues $\lambda_1=281.6890$, $\lambda_2=224.7816$, $\lambda_3=151.2445$, and $\lambda_4=127.9531$. The standard deviations $\mathrm{std}(\mathbf{v}_i|_D)$ within $D$ are $4.4695\times10^{-6}$, $8.0139\times10^{-3}$, $2.7979\times10^{-2}$, and $2.1694\times10^{-2}$, respectively. Therefore, the first principal eigenvector is selected to construct the mask $\mathbf{M}$ and the `mean prior' $\widetilde{\boldsymbol{\delta\sigma}}$.}
    \label{fig:3d_eigs_first4_short}
\end{figure}

For the proposed algorithm, we took the 50 dB case as an example to illustrate the construction process of $\mathbf{M}$ and $\widetilde{\boldsymbol{\delta\sigma}}$ in the local structural prior defined in Eq.~\eqref{eq:SGLSP}, as shown in Fig.~\ref{fig:3d_prior_roi_short}. We used the NWATV initial reconstruction $\boldsymbol{\delta\sigma}^\lambda$ as the reference image, as shown in Fig.~\ref{fig:3d_prior_roi_short}(a). First, we selected a problematic region $D$ away from $\partial\Omega$, as shown in Fig.~\ref{fig:3d_prior_roi_short}(b). Then, we substituted $\boldsymbol{\delta\sigma}^\lambda$ into Eq.~\eqref{eq:Gaussian_similarity_matrix} to compute the Gaussian similarity matrix $\mathbf{C}$, where we determined the kernel width parameter $h$ by the standard deviation of $\boldsymbol{\delta\sigma}^\lambda$ within $D$. In this experiment, $\mathrm{std}(\boldsymbol{\delta\sigma}^\lambda|_D)=0.019814$, and we set the empirical constant $\alpha$ to 1.56, giving $h=0.031$ from Eq.~\eqref{eq:kernel_width}. Subsequently, we eigendecomposed the constructed matrix $\mathbf{C}$ according to Eq.~\eqref{eq:eigendecomposition_C} and compared the first four principal eigenvectors in Fig.~\ref{fig:3d_eigs_first4_short}. According to the minimum-standard-deviation criterion in $D$ defined in Eq.~\eqref{eq:selected_eigenvector}, we selected the first principal eigenvector in this example, as shown in Fig.~\ref{fig:3d_prior_roi_short}(c). We then applied the threshold condition $\mathbf{v}_1>0.01$ to the selected eigenvector. We defined the finite elements satisfying this condition as the foreground region $\mathcal{F}$ and the remaining finite elements as the background region $\mathcal{B}$. We then obtained the binary mask $\mathbf{M}$ according to Eq.~\eqref{eq:segmentation_mask}, as shown in Fig.~\ref{fig:3d_prior_roi_short}(d). Then, we computed the mean of $\boldsymbol{\delta\sigma}^\lambda$ over the foreground region $\mathcal{F}$ and constructed the `mean prior' $\widetilde{\boldsymbol{\delta\sigma}}$ according to Eqs.~\eqref{eq:foreground_mean} and \eqref{eq:mean_prior}, as shown in Fig.~\ref{fig:3d_prior_roi_short}(e). Together, $\mathbf{M}$ and $\widetilde{\boldsymbol{\delta\sigma}}$ formed the local structural prior in Eq.~\eqref{eq:SGLSP}. Based on this constructed local structural prior, we solved the regularized data-fitting problem defined in \eqref{minimization:final_regularized}. We set the parameters of the proposed algorithm as $\beta=0.005$; at 50 dB, $\lambda=6\times10^{-14}$, $\rho=4\times10^{-12}$, $\lambda_1=4\times10^{-10}$, and $\lambda_2=4\times10^{-4}$; at 30 dB, $\lambda=2.5\times10^{-14}$, $\rho=2.5\times10^{-12}$, $\lambda_1=2.5\times10^{-10}$, and $\lambda_2=2.5\times10^{-4}$.

\begin{figure*}[!t]
    \centering
    \includegraphics[width=0.90\textwidth]{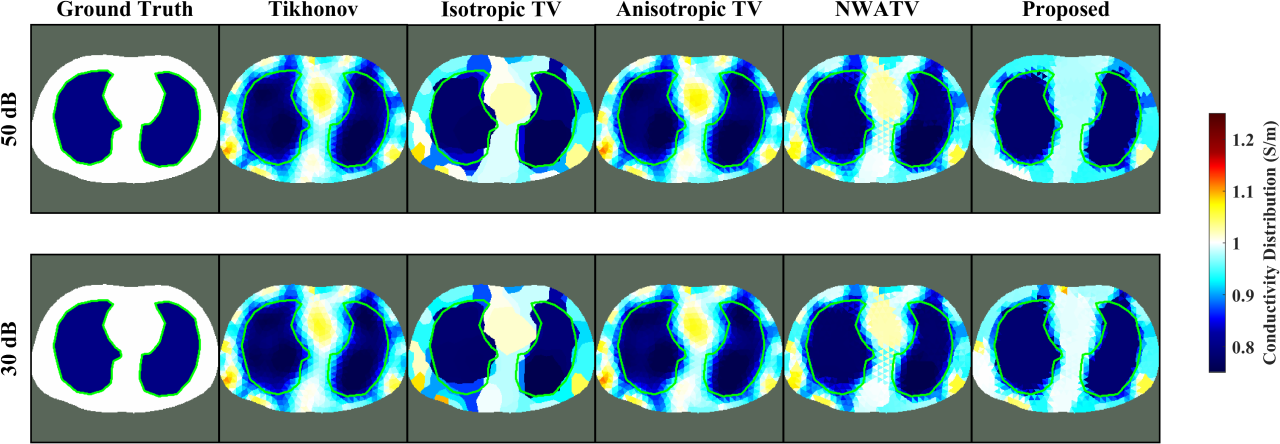}
    \caption{Comparison of two-dimensional absolute-conductivity reconstruction results obtained by different algorithms on the projection plane of the three-dimensional thorax model under 50 dB and 30 dB measurement-noise conditions.}
    \label{fig:3d_recon_short}
\end{figure*}

Fig. \ref{fig:3d_recon_short} shows the reconstruction results of the five algorithms on the central slice under different noise levels. Because the anatomical gap between the left and right lungs is narrow and the spatial sensitivity distribution is nonuniform in the three-dimensional thorax model, conventional reconstruction methods tend to produce target merging and boundary blurring in regions with relatively low spatial resolution. By contrast, the proposed method better preserves the spatial separability of the left and right lungs under both 50 dB and 30 dB conditions, mainly improving the target-merging problem in spatially ambiguous regions while also reducing local artifacts near the electrodes to some extent.

\begin{figure}[htbp]
    \centering
    \begin{minipage}{0.48\columnwidth}
        \centering
        \includegraphics[width=\linewidth]{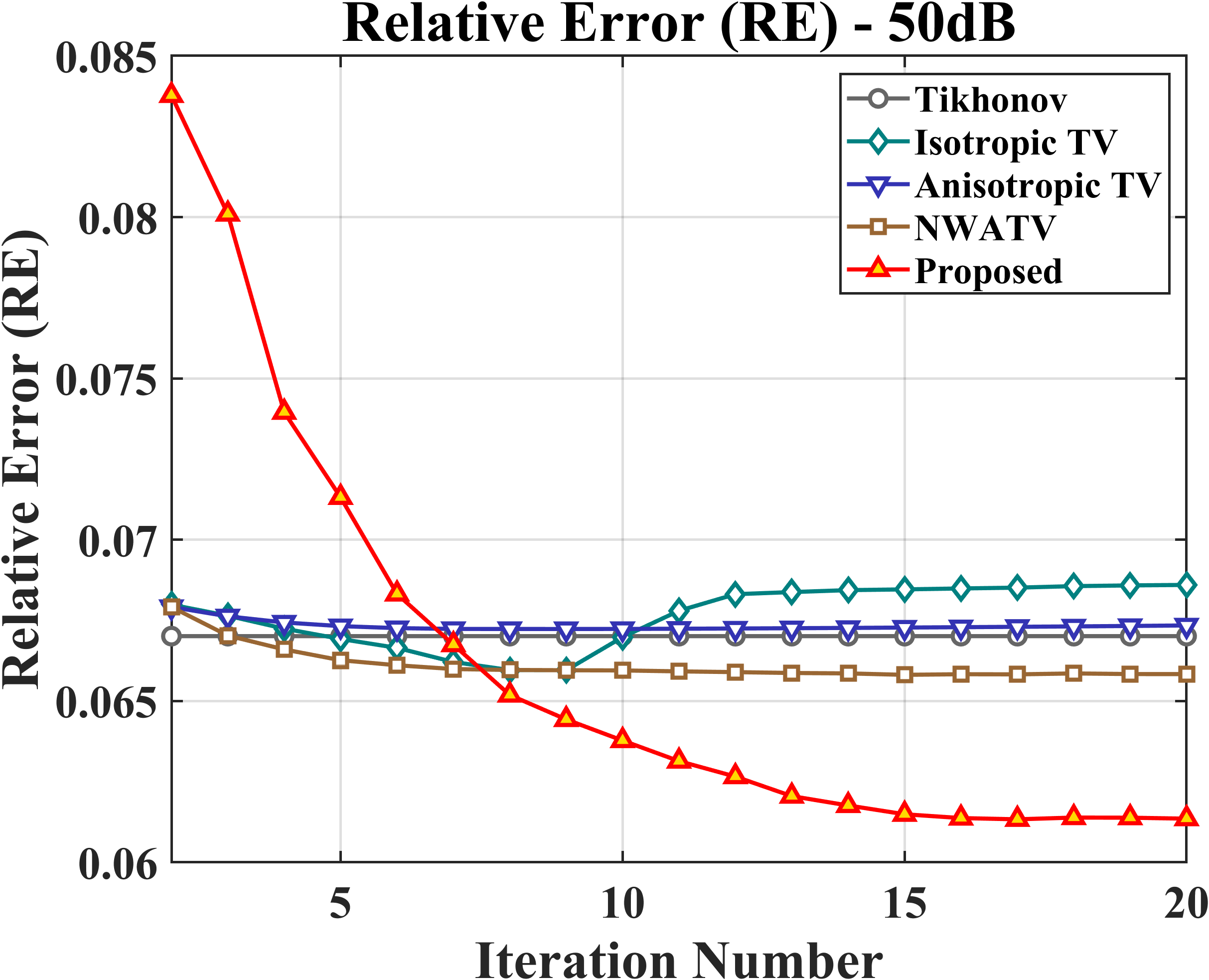}
        \centerline{\footnotesize (a) RE (50 dB)}
    \end{minipage}\hfill
    \begin{minipage}{0.48\columnwidth}
        \centering
        \includegraphics[width=\linewidth]{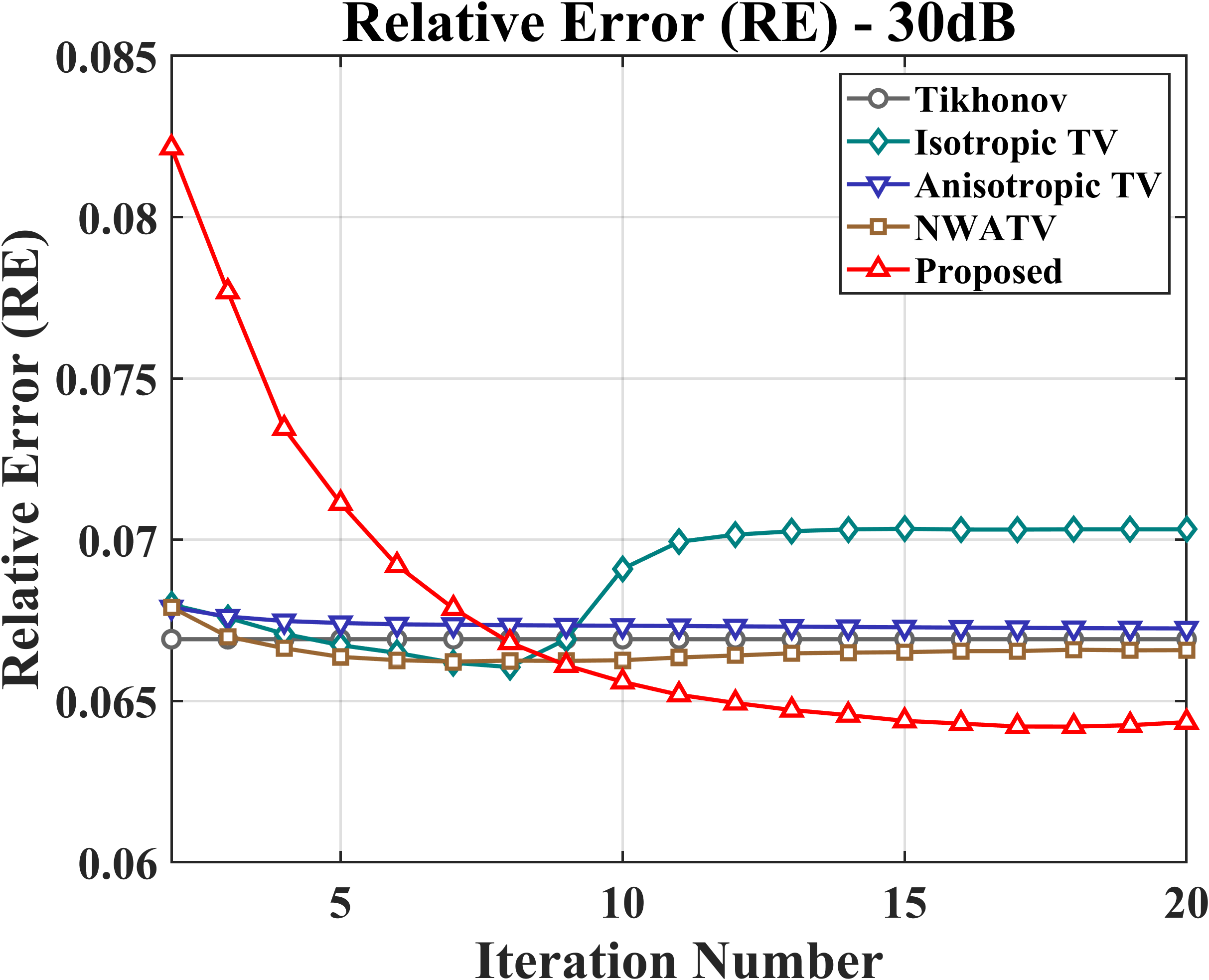}
        \centerline{\footnotesize (b) RE (30 dB)}
    \end{minipage}

    \vspace{0.1cm}

    \begin{minipage}{0.48\columnwidth}
        \centering
        \includegraphics[width=\linewidth]{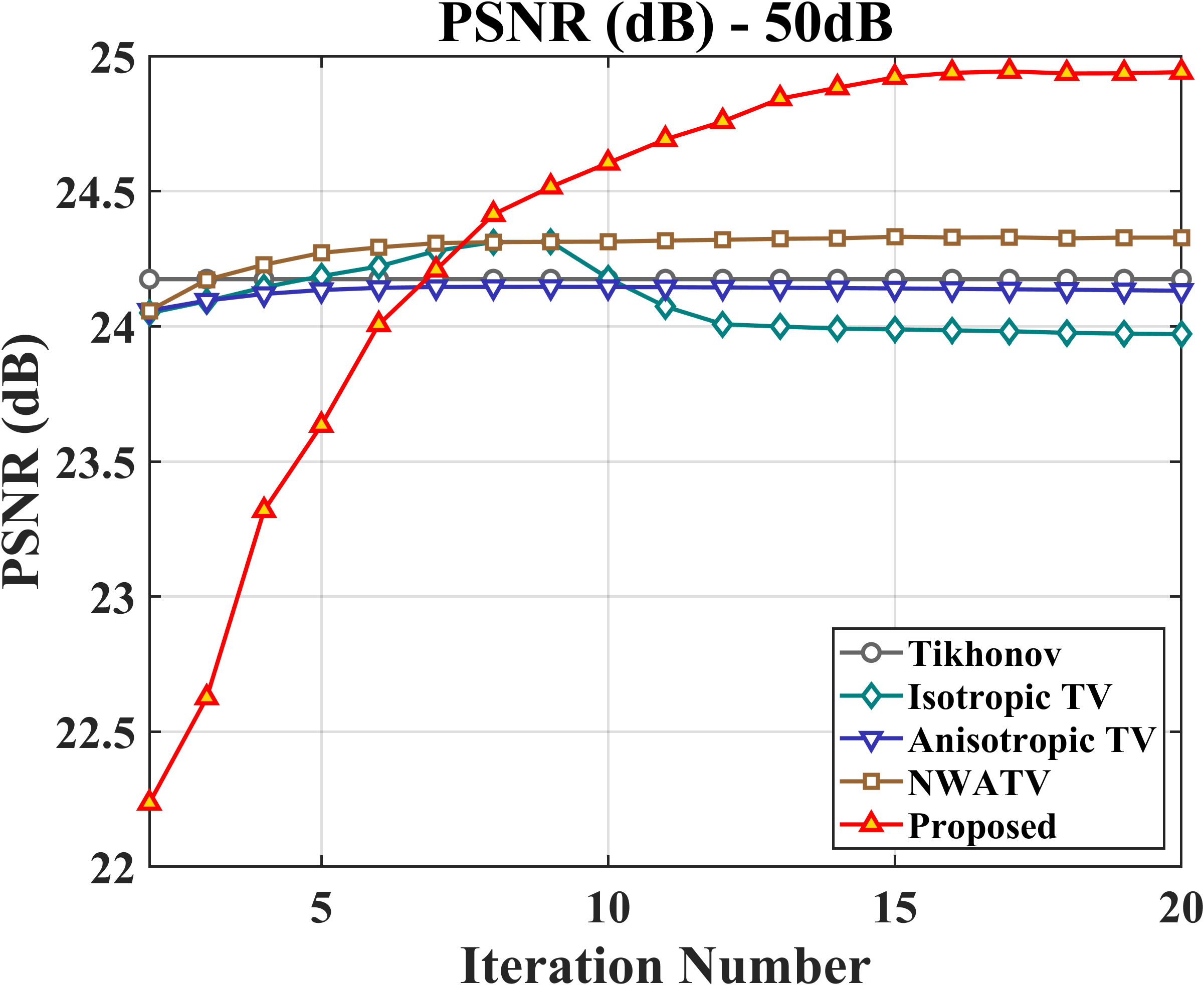}
        \centerline{\footnotesize (c) PSNR (50 dB)}
    \end{minipage}\hfill
    \begin{minipage}{0.48\columnwidth}
        \centering
        \includegraphics[width=\linewidth]{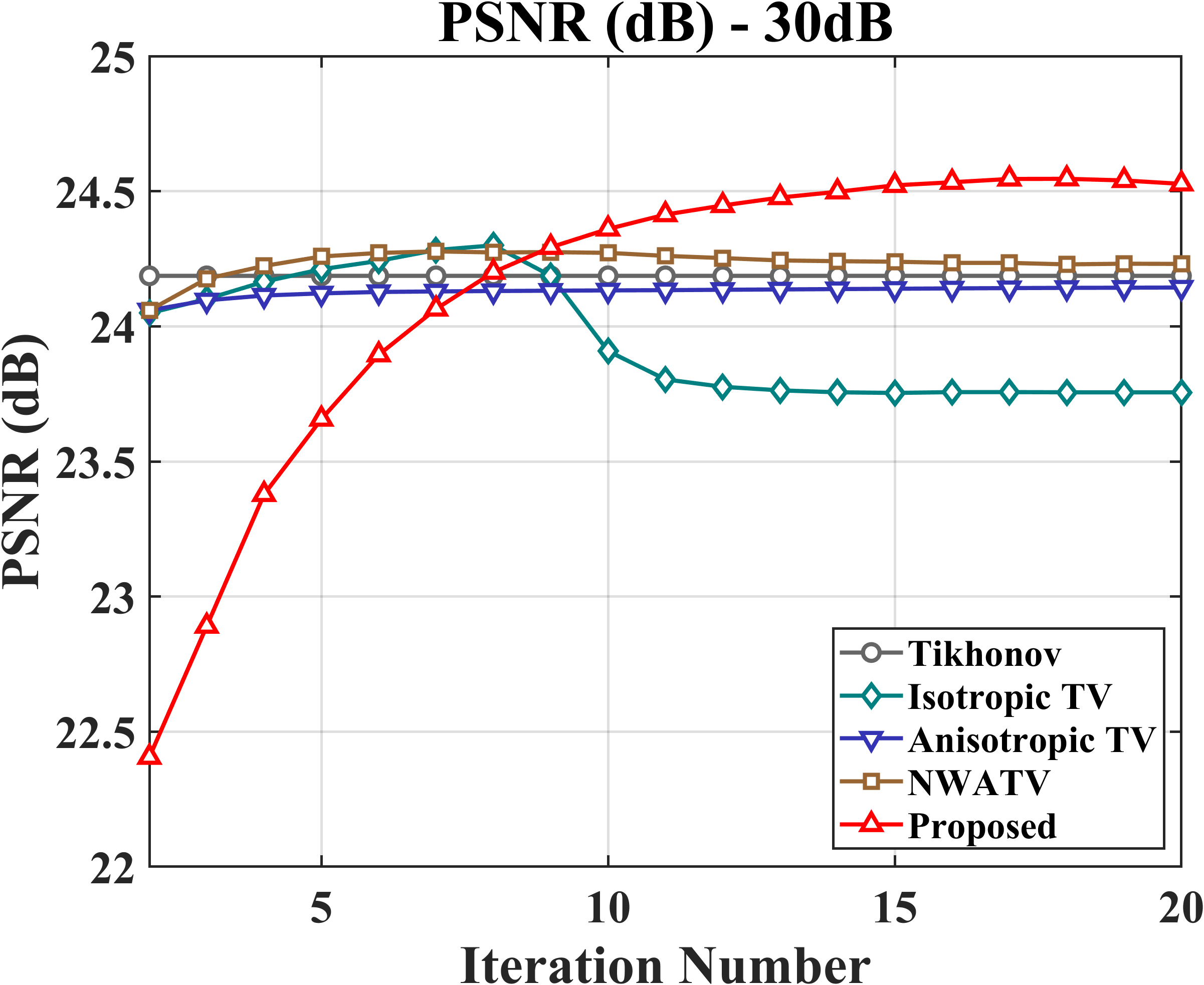}
        \centerline{\footnotesize (d) PSNR (30 dB)}
    \end{minipage}

    \vspace{0.1cm}

    \begin{minipage}{0.48\columnwidth}
        \centering
        \includegraphics[width=\linewidth]{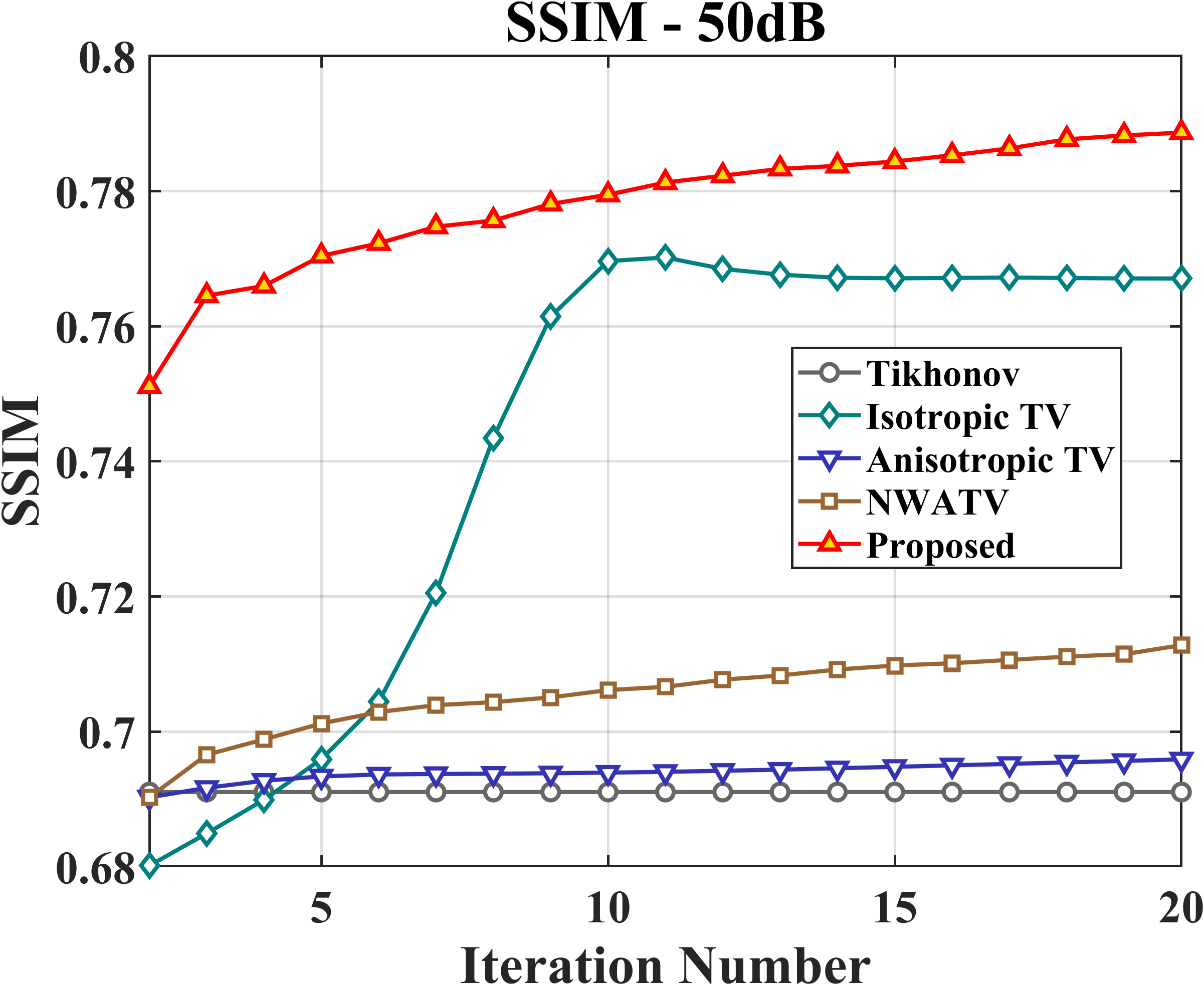}
        \centerline{\footnotesize (e) SSIM (50 dB)}
    \end{minipage}\hfill
    \begin{minipage}{0.48\columnwidth}
        \centering
        \includegraphics[width=\linewidth]{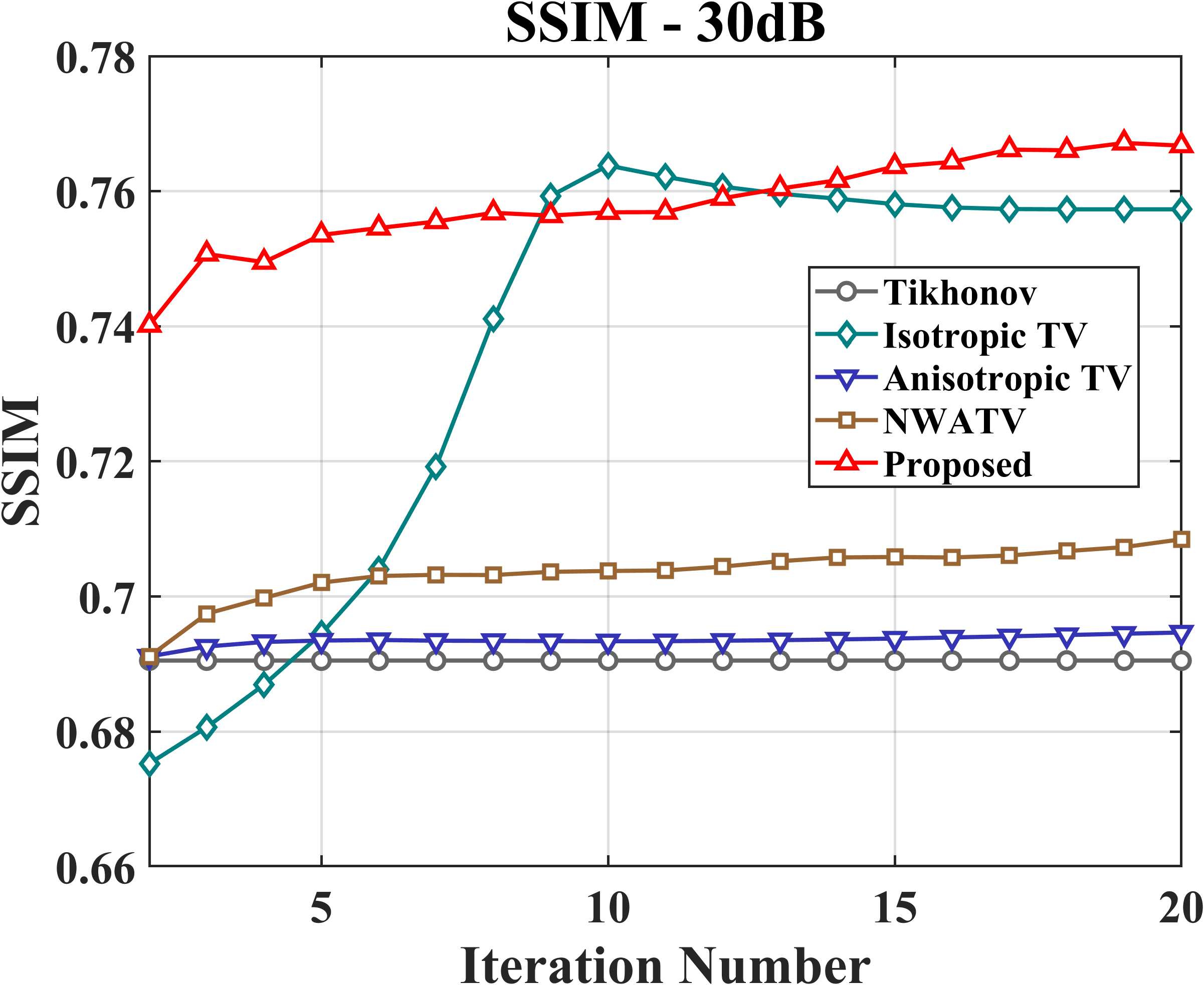}
        \centerline{\footnotesize (f) SSIM (30 dB)}
    \end{minipage}
    \caption{Comparison of the evolution curves of relative error (RE), peak signal-to-noise ratio (PSNR), and structural similarity index (SSIM) with respect to the iteration steps in the simulation experiment.}
    \label{fig:3d_metrics_short}
\end{figure}

Fig.~\ref{fig:3d_metrics_short} shows the evolution curves of RE, PSNR, and SSIM with respect to the iteration steps under the 50 dB and 30 dB noise levels. Overall, the proposed method yields lower RE and higher PSNR and SSIM values than the comparison methods. Table~\ref{tab:3d_metrics_noise} gives the final quantitative metrics corresponding to Fig.~\ref{fig:3d_metrics_short}. The proposed method achieves the lowest RE and the highest PSNR and SSIM under both noise levels, indicating improved reconstruction accuracy and robustness in the three-dimensional thorax simulation.

\begin{table*}[!t]
    \centering
    \caption{Quantitative comparison of reconstruction metrics under different noise levels in the 3-D thorax simulation.}
    \label{tab:3d_metrics_noise}
    \renewcommand{\arraystretch}{1.2}
    \scriptsize
    \begin{tabular}{lcccccc}
        \toprule
        \textbf{Algorithm} & \multicolumn{3}{c}{\textbf{50 dB}} & \multicolumn{3}{c}{\textbf{30 dB}} \\
        \cmidrule(lr){2-4} \cmidrule(lr){5-7}
        & \textbf{RE} & \textbf{PSNR (dB)} & \textbf{SSIM} & \textbf{RE} & \textbf{PSNR (dB)} & \textbf{SSIM} \\
        \midrule
        Tikhonov & 0.0670 & 24.17 & 0.6911 & 0.0669 & 24.19 & 0.6905 \\
        Isotropic TV & 0.0686 & 23.97 & 0.7671 & 0.0703 & 23.76 & 0.7573 \\
        Anisotropic TV & 0.0673 & 24.13 & 0.6959 & 0.0673 & 24.14 & 0.6947 \\
        NWATV & 0.0658 & 24.33 & 0.7128 & 0.0666 & 24.23 & 0.7085 \\
        Proposed & \textbf{0.0614} & \textbf{24.94} & \textbf{0.7887} & \textbf{0.0643} & \textbf{24.53} & \textbf{0.7668} \\
        \bottomrule
    \end{tabular}
\end{table*}

\subsection{Water Tank Experiments}

\begin{figure}[!t]
    \centering
    \begin{minipage}{0.40\columnwidth}
        \centering
        \includegraphics[width=\linewidth]{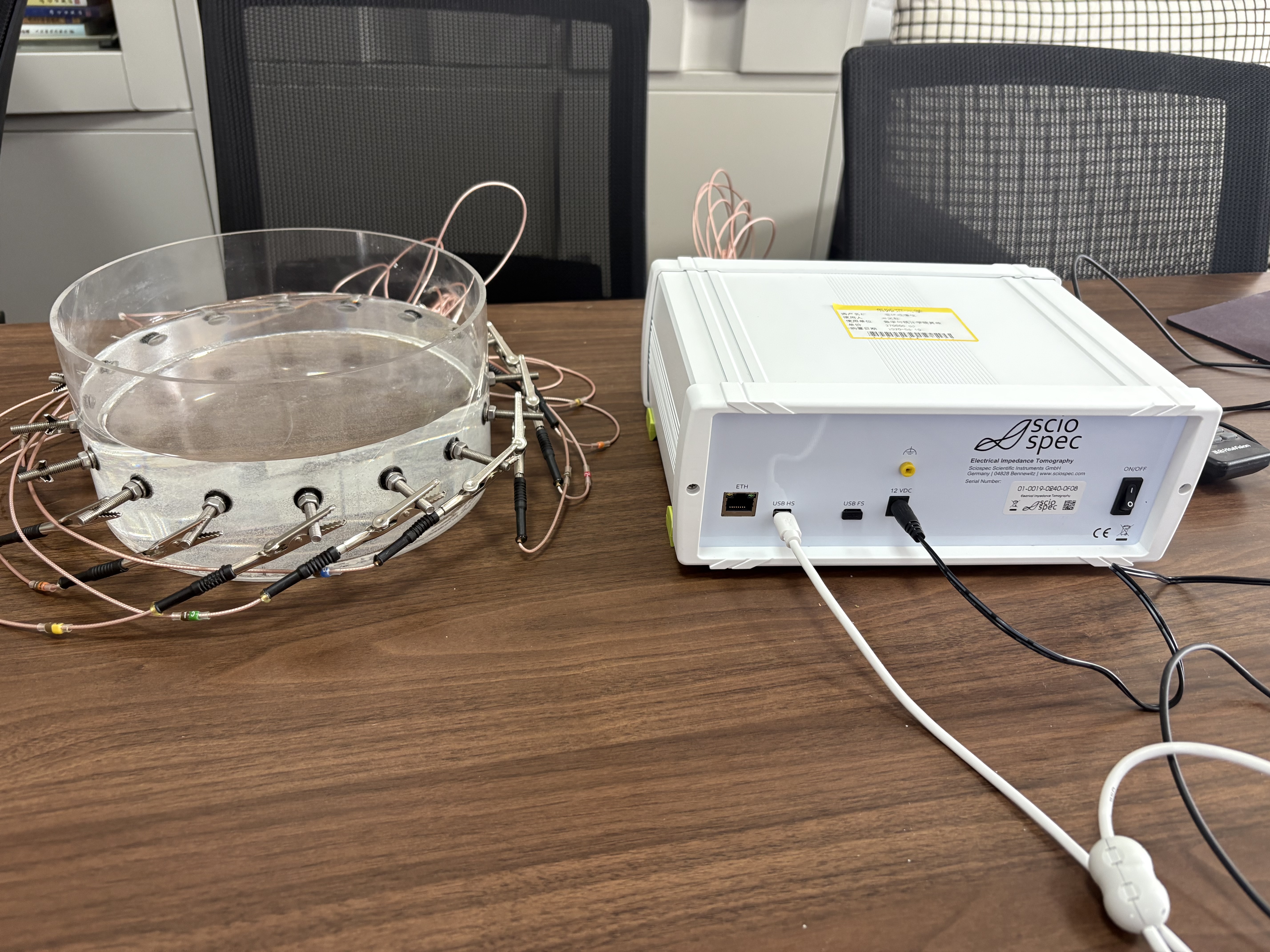}
        \centerline{\footnotesize (a)}
    \end{minipage}
    \hspace{0.16\columnwidth}
    \begin{minipage}{0.30\columnwidth}
        \centering
        \includegraphics[width=\linewidth]{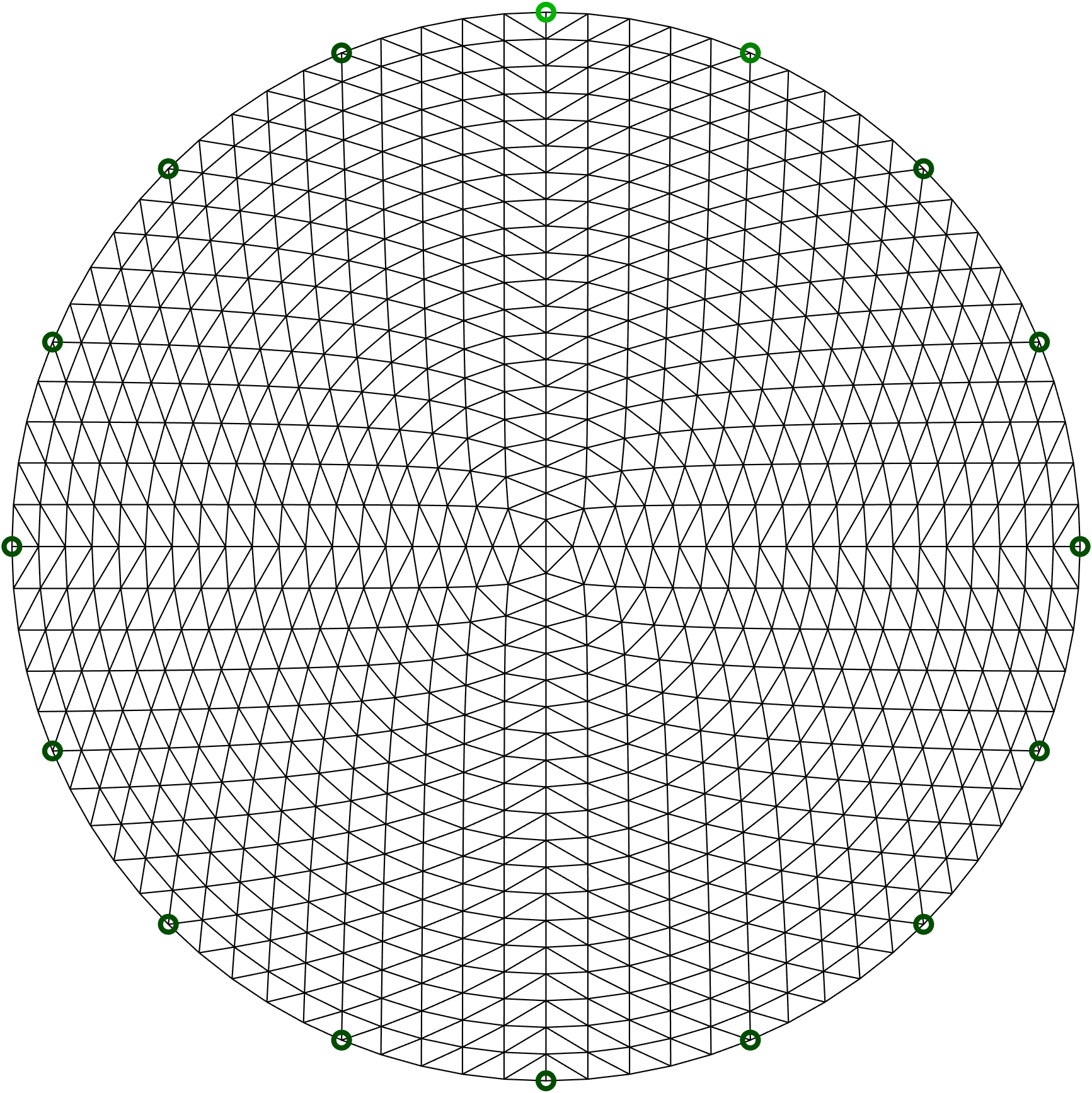}
        \centerline{\footnotesize (b)}
    \end{minipage}

    \vspace{0.6em}

    \begin{minipage}{0.40\columnwidth}
        \centering
        \includegraphics[width=\linewidth]{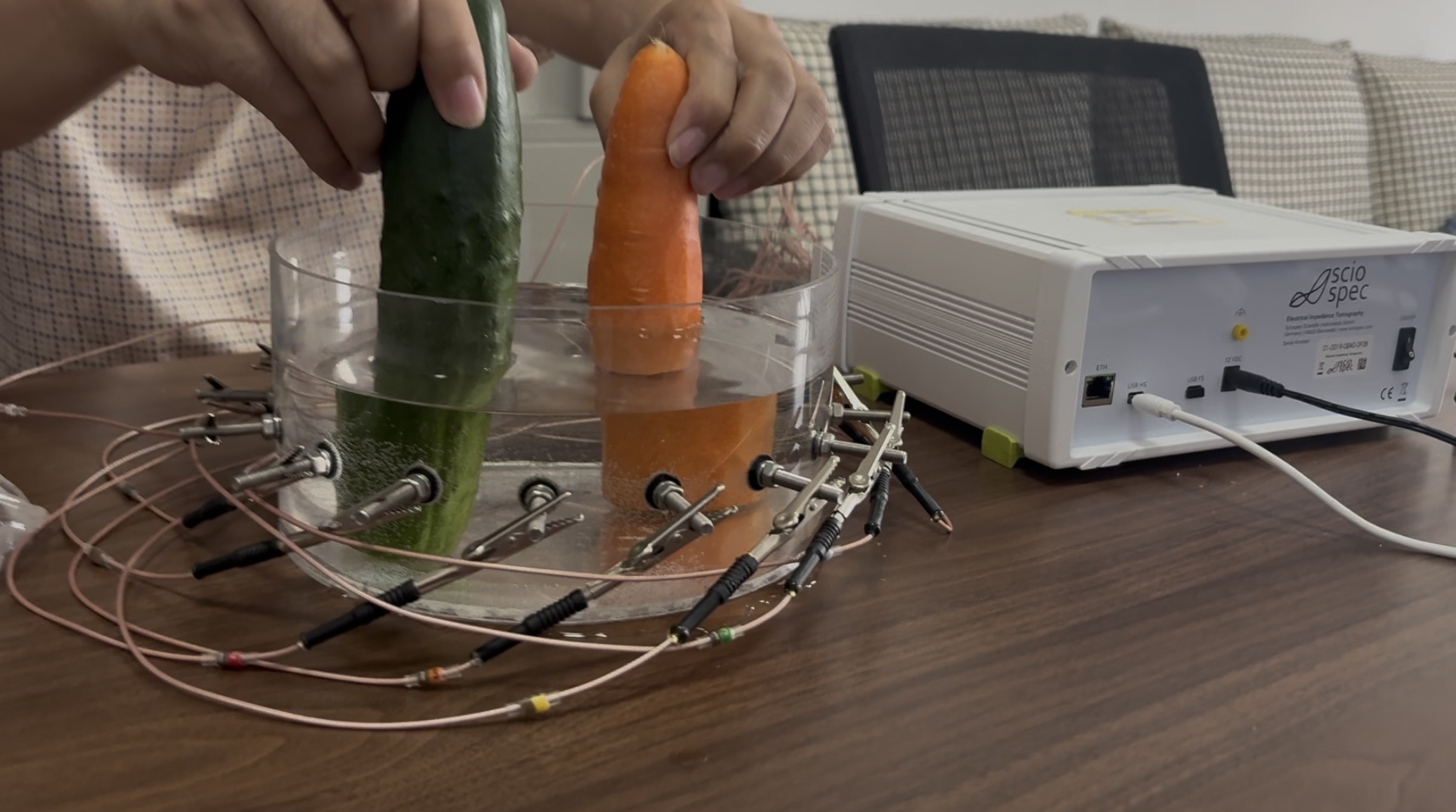}
        \centerline{\footnotesize (c)}
    \end{minipage}
    \hspace{0.04\columnwidth}
    \begin{minipage}{0.42\columnwidth}
        \centering
        \includegraphics[width=\linewidth]{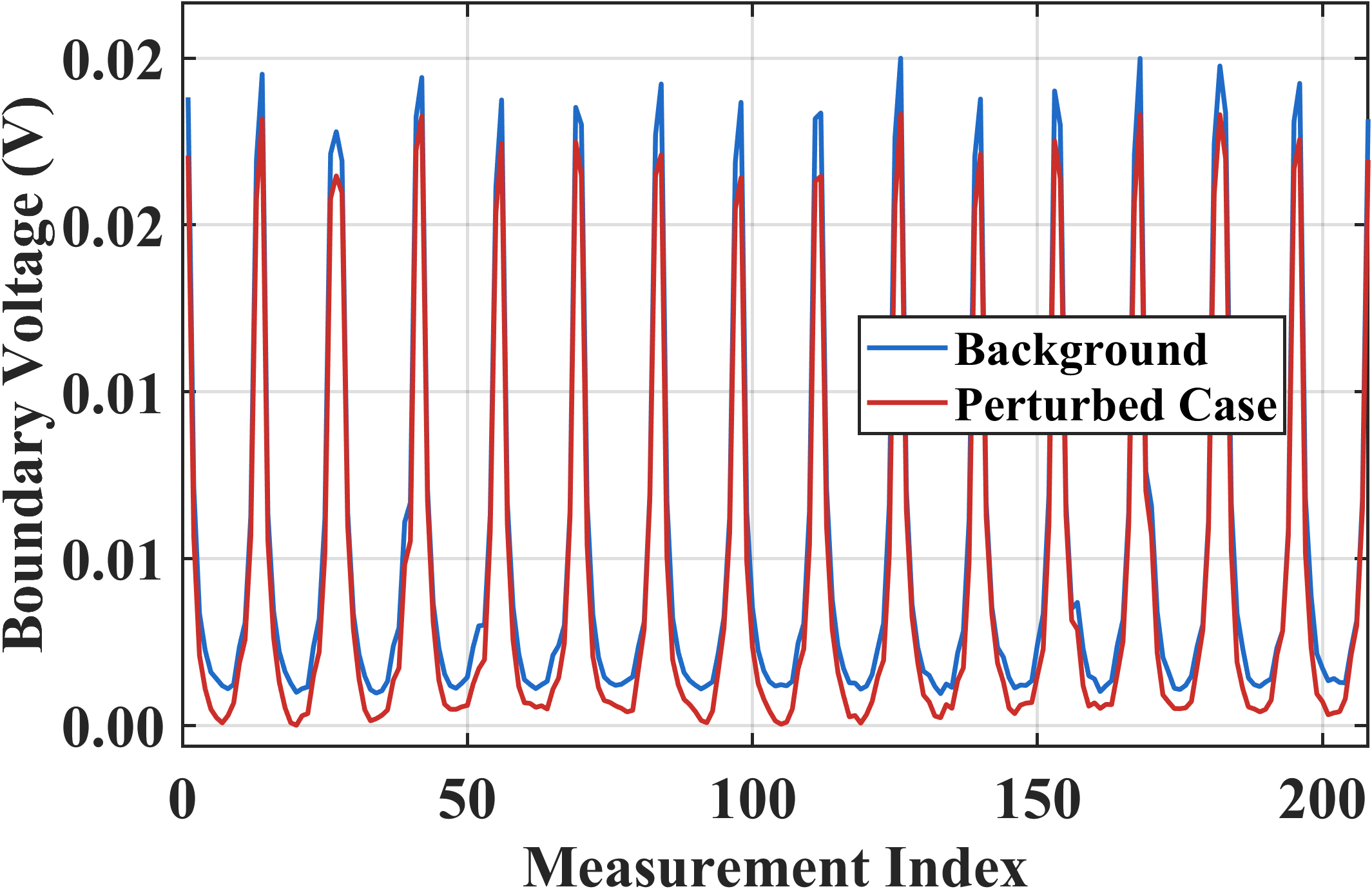}
        \centerline{\footnotesize (d)}
    \end{minipage}
    \caption{Water tank experimental setup, finite element model, and measured boundary voltages. (a) Cylindrical tank apparatus in the water tank experiment. (b) Two-dimensional finite element mesh used for inverse reconstruction. (c) Photograph of the carrot-cucumber two-target tank experiment. (d) Measured boundary voltage vectors under the background and target-perturbed states in the carrot-cucumber experiment, in which the blue and the red lines are the voltage vectors corresponding to $\sigma_0$ and $\sigma$, respectively.}
    \label{fig:tank_setup}
\end{figure}

To verify the capacity of the proposed method for experimentally measured data, we used a Sciospec 16-channel EIT system \cite{EIT} and a cylindrical experimental water tank to conduct a carrot-cucumber two-target water tank experiment. The cylindrical tank and electrode arrangement are shown in Fig.~\ref{fig:tank_setup}(a). We filled the tank with tap water as the homogeneous background medium; its inner diameter measured by a vernier caliper is $19.35\,\mathrm{cm}$. We uniformly placed sixteen electrodes around the tank. We set the imaging plane to the electrode plane, and the two-dimensional finite element reconstruction model contains 1600 triangular elements and 841 nodes, as shown in Fig.~\ref{fig:tank_setup}(b). The actual tank experiment scene is shown in Fig.~\ref{fig:tank_setup}(c). In the experiment, we injected a current with amplitude $1$ mA and frequency $100$ kHz through adjacent electrodes and measured the boundary voltages between adjacent electrode pairs. The data acquisition speed of the device is approximately 20 frames/s. Each frame contains $16\times13=208$ effective boundary voltage measurements for reconstruction. The background conductivity of water can be estimated from the boundary voltage data measured when no object is placed in the tank; we used $\sigma_0\approx4.67\times10^{-3}$ S/m \cite{adler2006eidors}. We computed the sensitivity matrix $\mathbf{S}$ from \eqref{eq:sensitivity_matrix} under the estimated $\sigma_0$.

\begin{figure*}[htbp]
    \centering
    \begin{minipage}{0.18\textwidth}
        \centering
        \includegraphics[width=\linewidth]{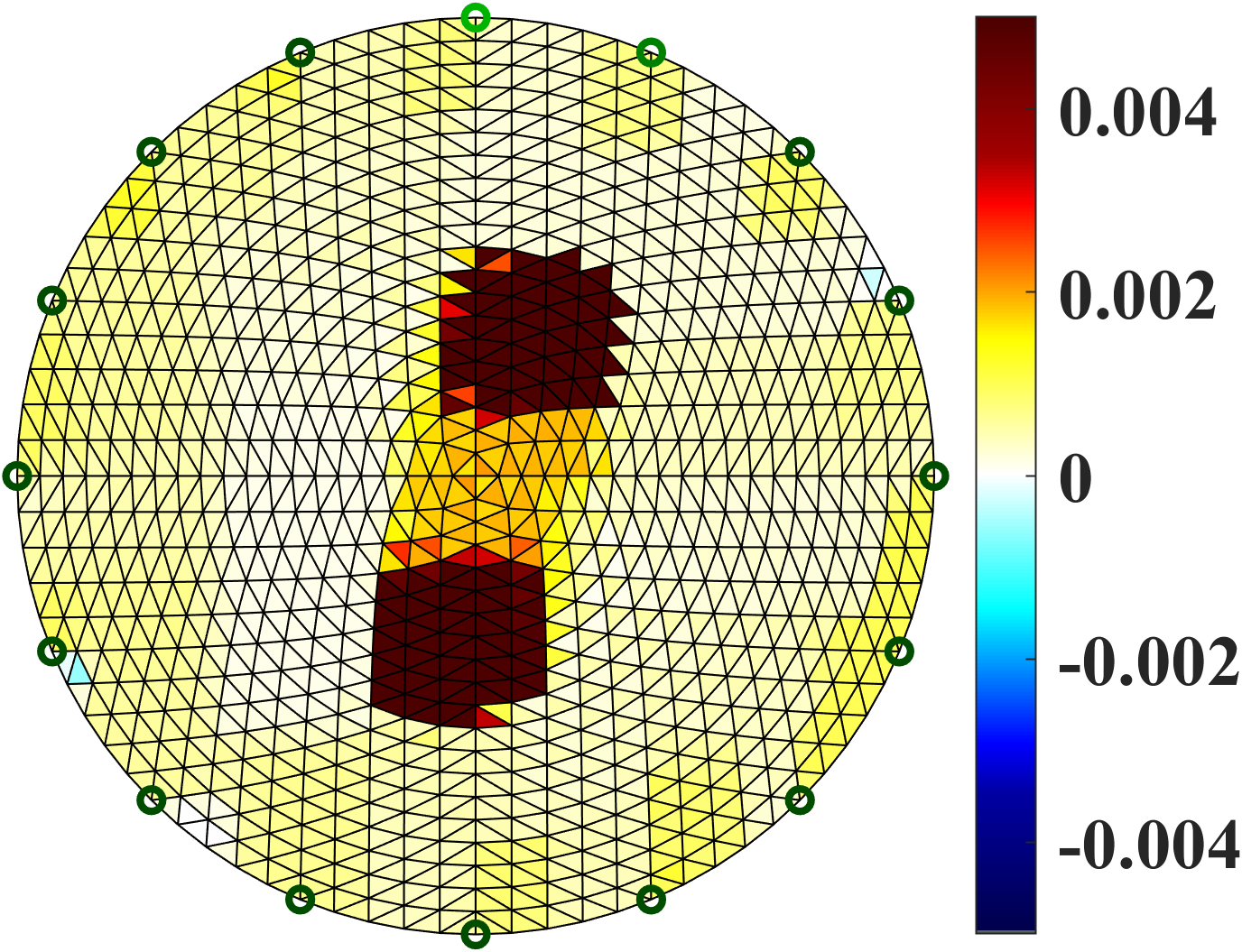}
        \centerline{\footnotesize (a)}
    \end{minipage}
    \hspace{0.01\textwidth}
    \begin{minipage}{0.18\textwidth}
        \centering
        \includegraphics[width=\linewidth]{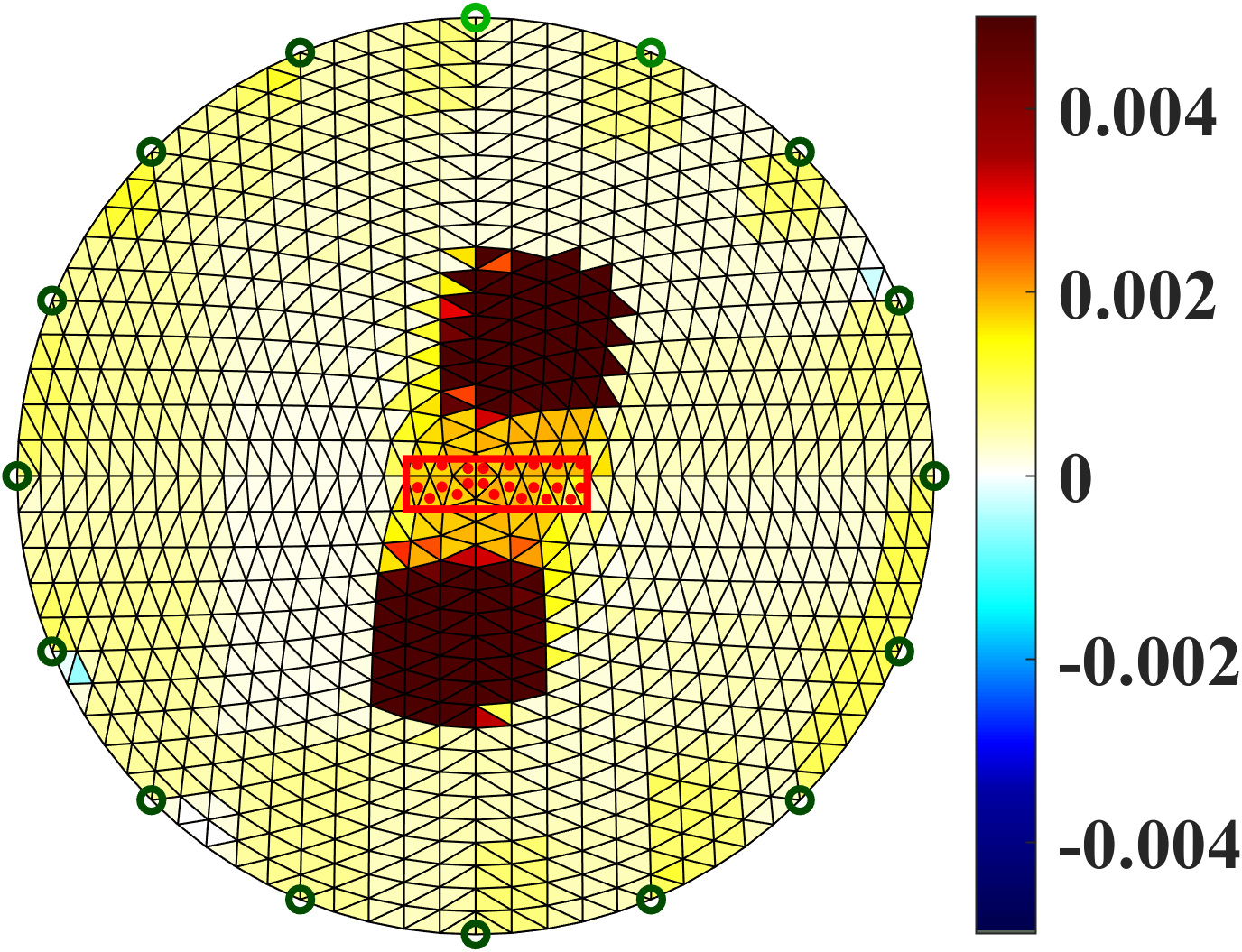}
        \centerline{\footnotesize (b)}
    \end{minipage}
    \hspace{0.01\textwidth}
    \begin{minipage}{0.18\textwidth}
        \centering
        \includegraphics[width=\linewidth]{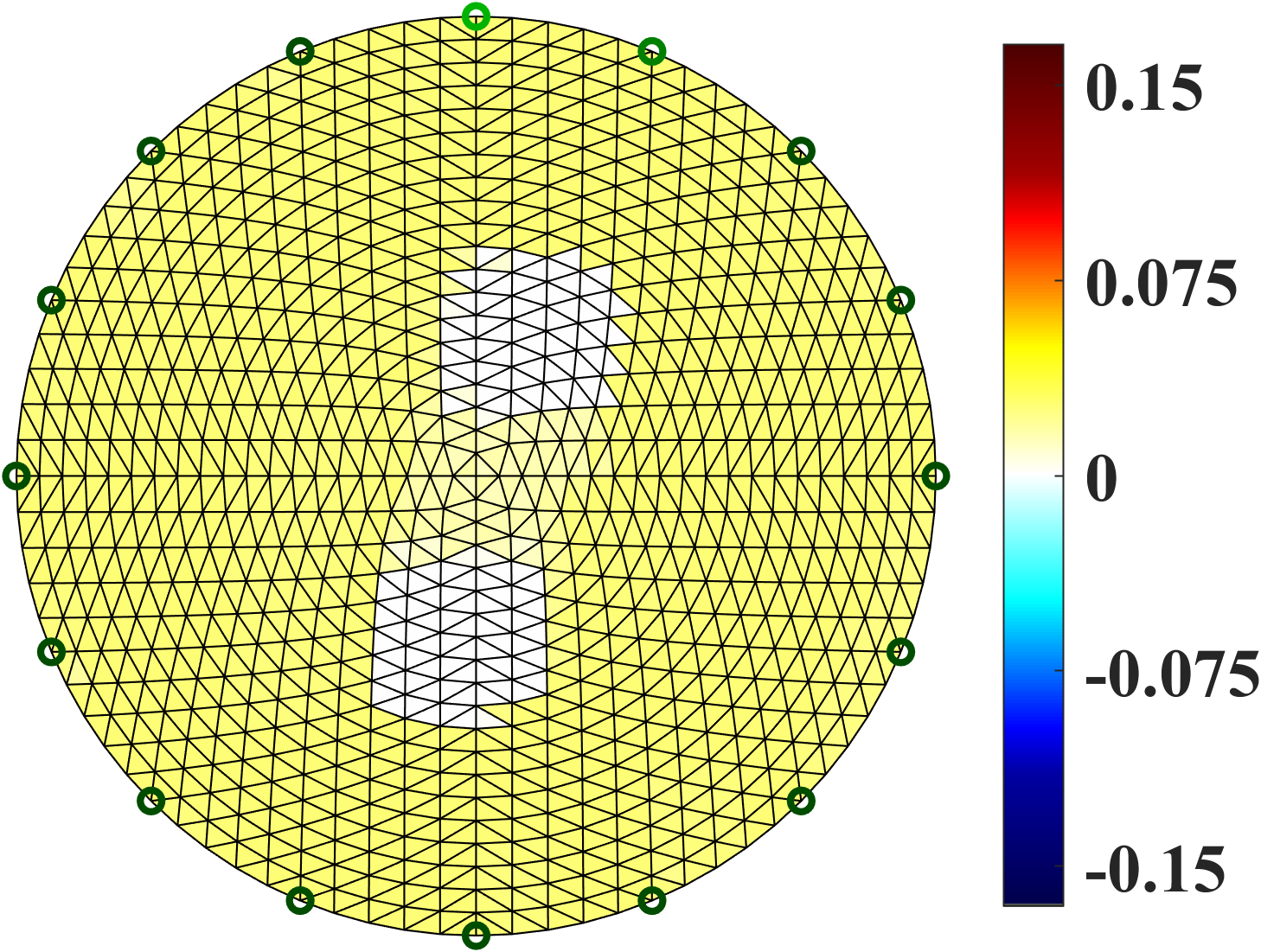}
        \centerline{\footnotesize (c)}
    \end{minipage}
    \hspace{0.01\textwidth}
    \begin{minipage}{0.14\textwidth}
        \centering
        \includegraphics[width=\linewidth]{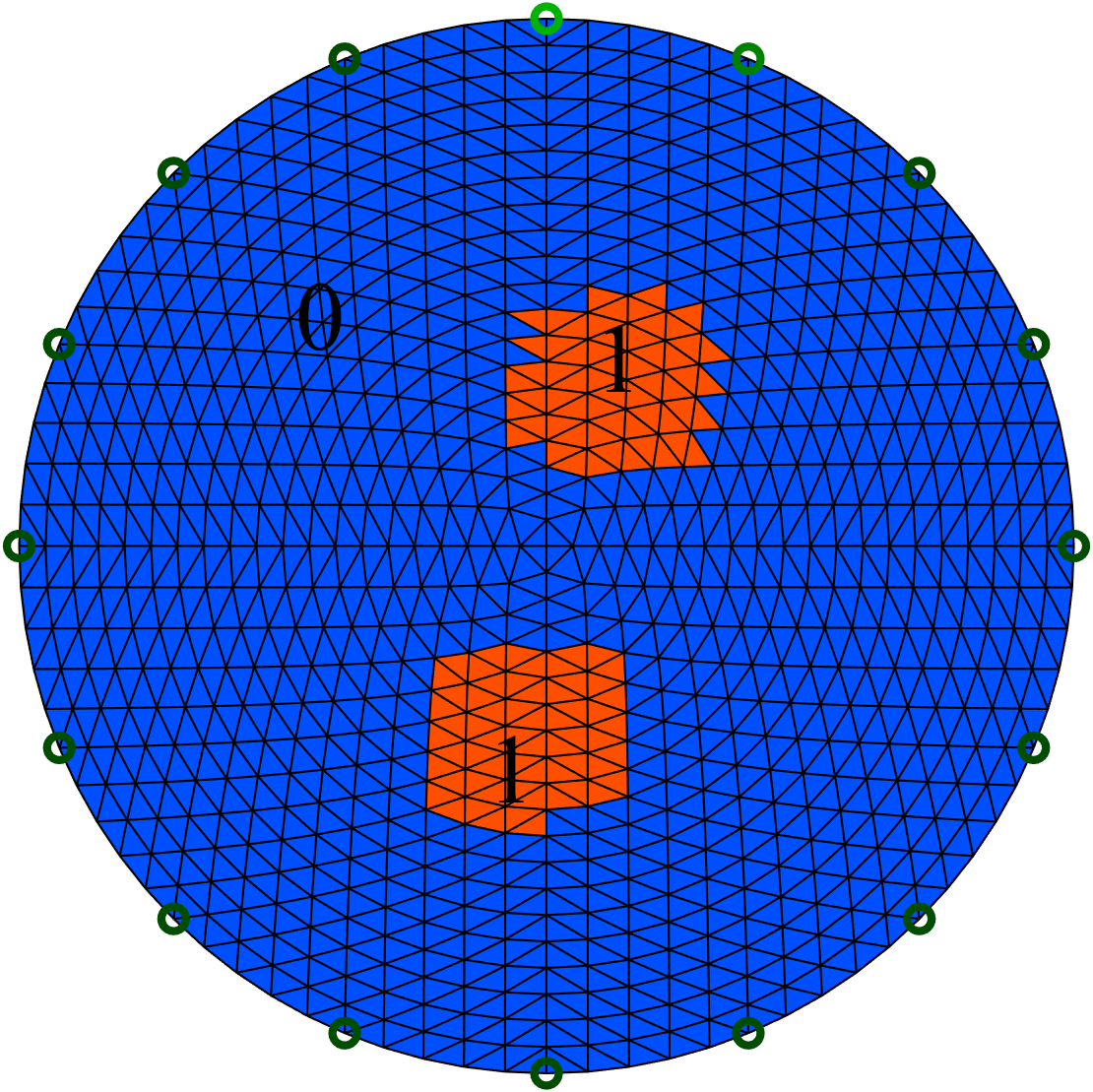}
        \centerline{\footnotesize (d)}
    \end{minipage}
    \hspace{0.01\textwidth}
    \begin{minipage}{0.18\textwidth}
        \centering
        \includegraphics[width=\linewidth]{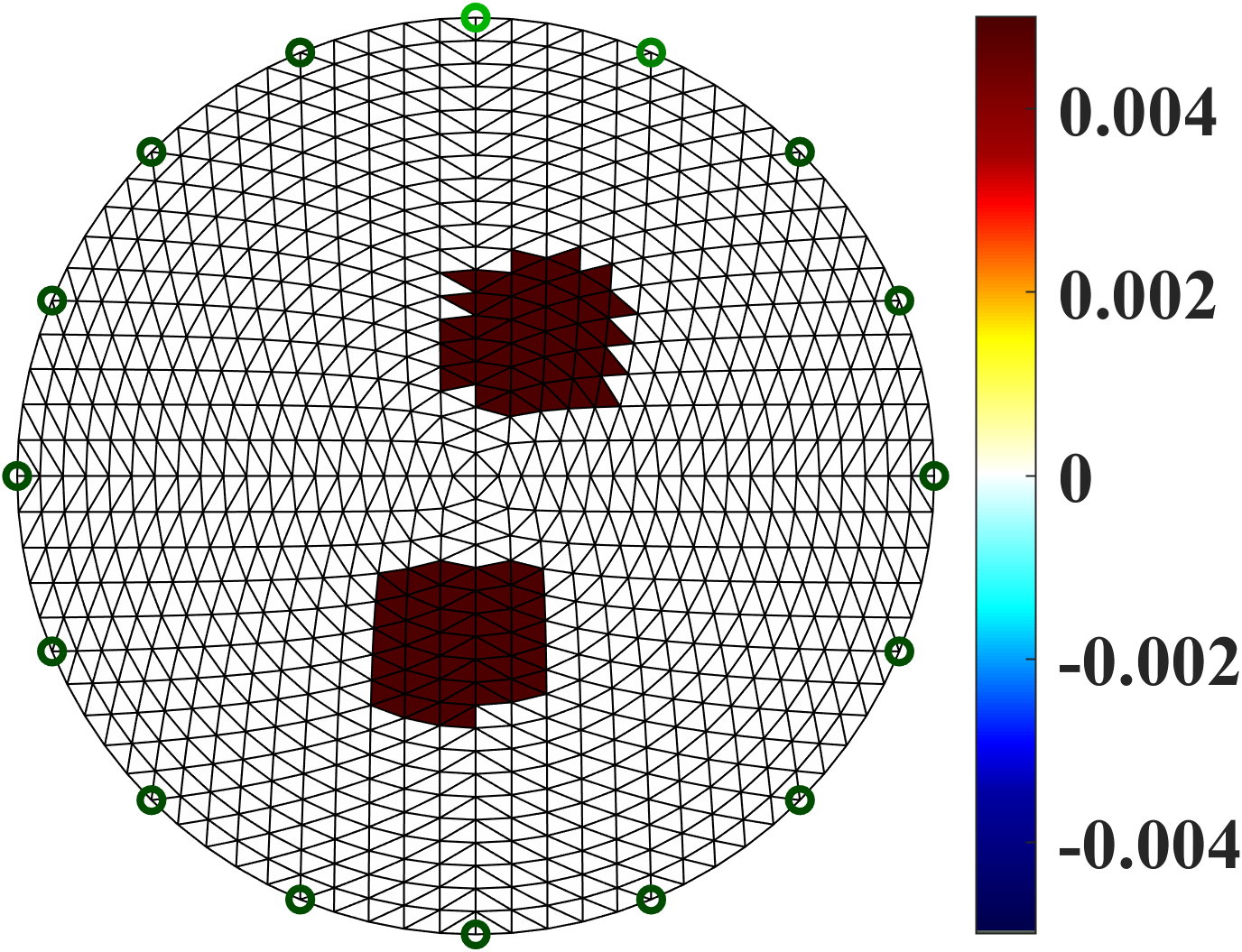}
        \centerline{\footnotesize (e)}
    \end{minipage}
    \caption{Extraction process of $\mathbf{M}$ and $\widetilde{\boldsymbol{\delta\sigma}}$ in the carrot-cucumber tank experiment, shown from (a) to (e). (a) Initial conductivity perturbation $\boldsymbol{\delta\sigma}^\lambda$ obtained from NWATV reconstruction. (b) Problematic region $D$ away from $\partial\Omega$. (c) Principal eigenvector of matrix $\mathbf{C}$ with the smallest standard deviation in $D$, namely the first principal eigenvector. (d) Mask $\mathbf{M}$ obtained by binary segmentation of the first principal eigenvector. (e) Assigned `mean prior' $\widetilde{\boldsymbol{\delta\sigma}}$.}
    \label{fig:tank_prior_short}
\end{figure*}

\begin{figure}[htbp]
    \centering
    \begin{minipage}{0.35\columnwidth}
        \centering
        \includegraphics[width=\linewidth]{figures/fig_27_figure4b_eigen1_unifiedcolorbar_algo3.png}
        \centerline{\footnotesize (a)}
    \end{minipage}\hspace{0.06\columnwidth}
    \begin{minipage}{0.35\columnwidth}
        \centering
        \includegraphics[width=\linewidth]{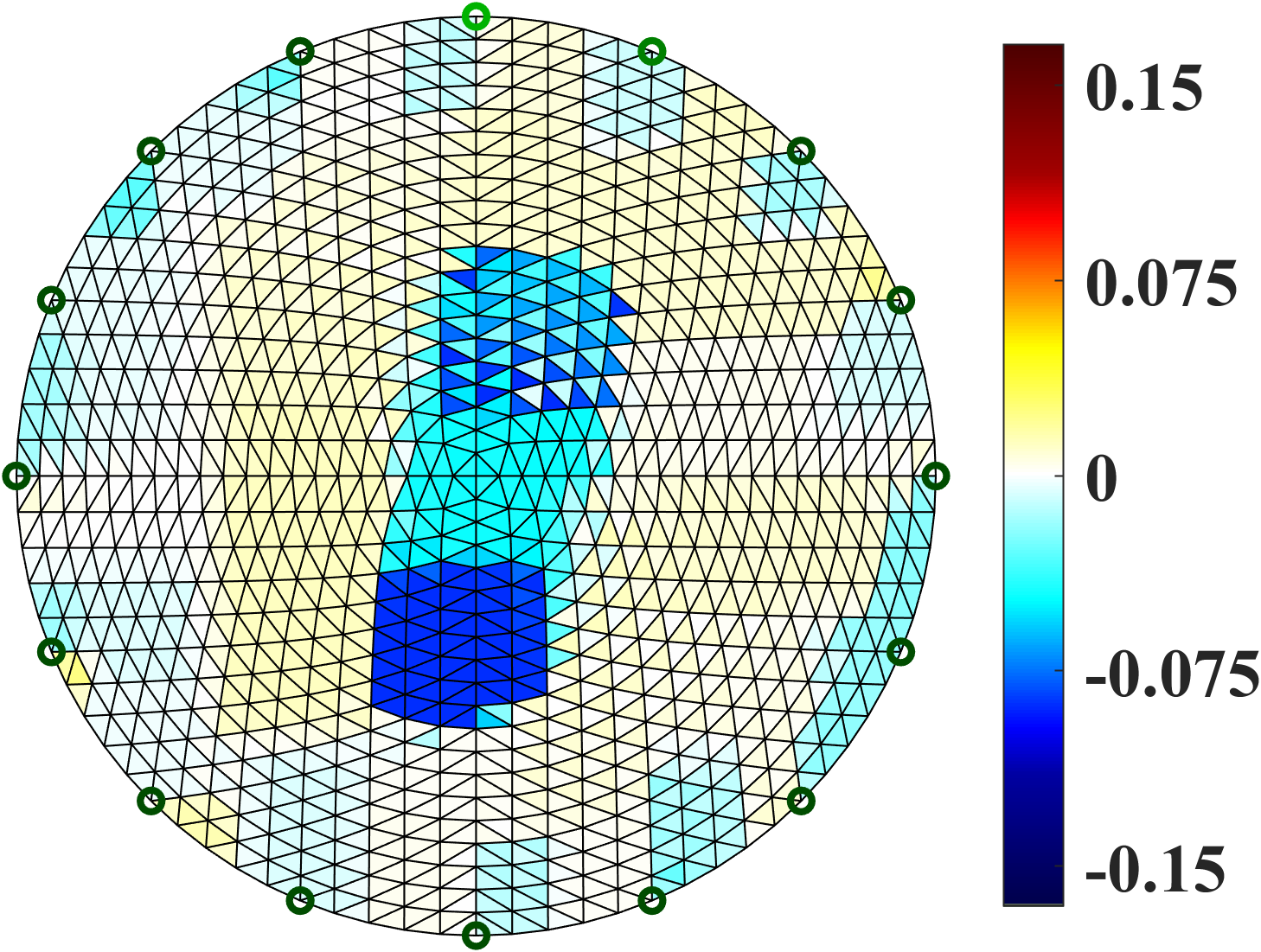}
        \centerline{\footnotesize (b)}
    \end{minipage}

    \vspace{0.2em}

    \begin{minipage}{0.35\columnwidth}
        \centering
        \includegraphics[width=\linewidth]{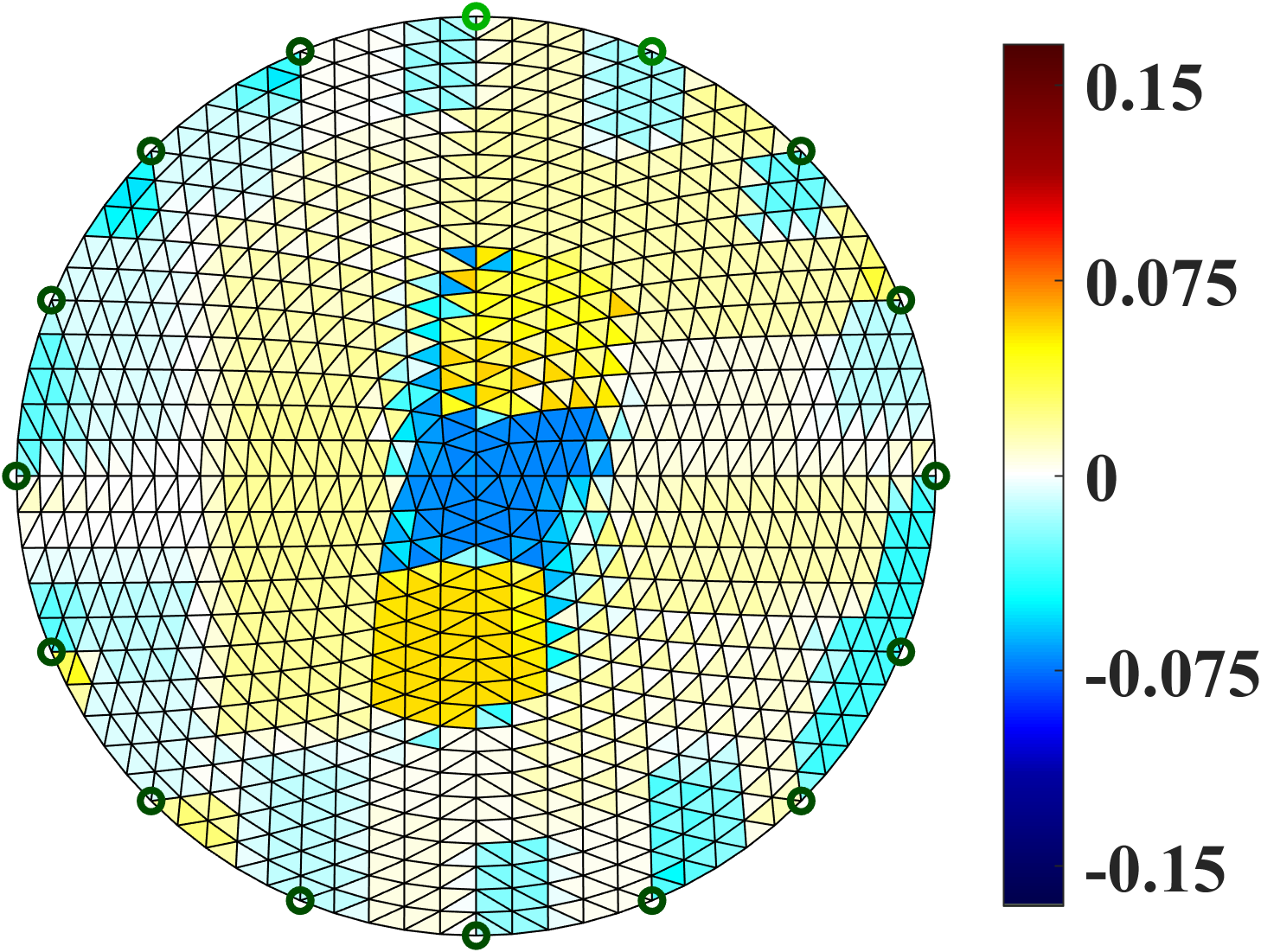}
        \centerline{\footnotesize (c)}
    \end{minipage}\hspace{0.06\columnwidth}
    \begin{minipage}{0.35\columnwidth}
        \centering
        \includegraphics[width=\linewidth]{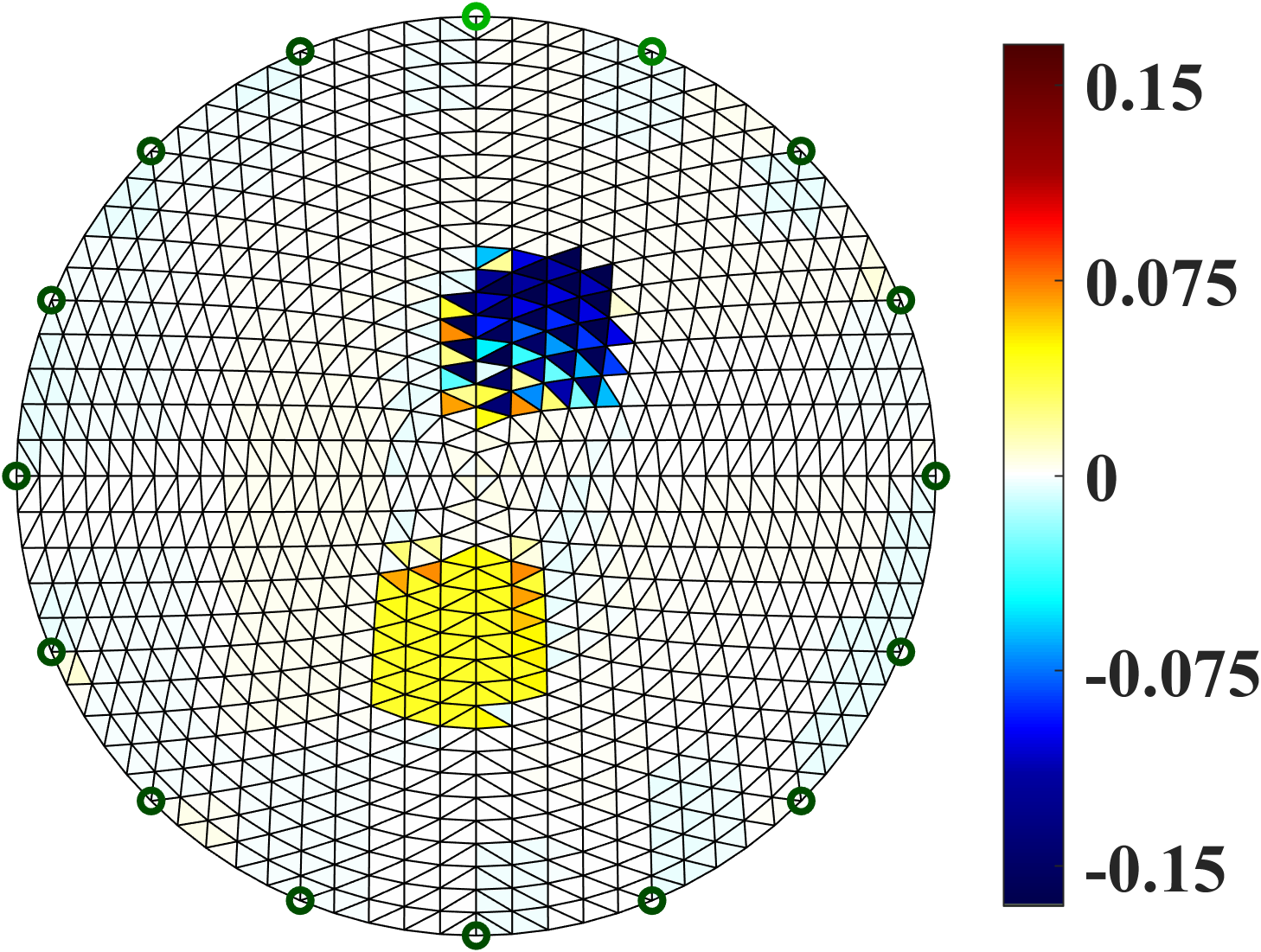}
        \centerline{\footnotesize (d)}
    \end{minipage}
    \caption{Comparison of the first four principal eigenvectors of the Gaussian similarity matrix $\mathbf{C}$ in the carrot--cucumber tank experiment. (a)-(d) show the first to fourth principal eigenvectors, respectively, with corresponding eigenvalues $\lambda_1=1.3807\times10^{3}$, $\lambda_2=8.7043\times10^{1}$, $\lambda_3=7.9375\times10^{1}$, and $\lambda_4=2.8708\times10^{1}$. The standard deviations $\mathrm{std}(\mathbf{v}_i|_D)$ within $D$ are $2.8640\times10^{-3}$, $7.2988\times10^{-3}$, $9.8321\times10^{-3}$, and $2.9077\times10^{-3}$, respectively. Therefore, the first principal eigenvector is selected to construct the mask $\mathbf{M}$ and the `mean prior' $\widetilde{\boldsymbol{\delta\sigma}}$.}
    \label{fig:tank_eigs_first4_short}
\end{figure}

We collected boundary voltage data under the homogeneous background state and the target-perturbed state, as shown in Fig.~\ref{fig:tank_setup}(d). Their difference yielded the measured voltage-difference data. We used the measured voltage-difference data as the reconstruction input and applied the proposed algorithm and four comparison methods to reconstruct the conductivity images. For the proposed algorithm, we followed the same prior-construction procedure as in the three-dimensional numerical experiment. We constructed the mask $\mathbf{M}$ and the `mean prior' $\widetilde{\boldsymbol{\delta\sigma}}$ in the local structural prior \eqref{eq:SGLSP} from the NWATV reconstruction $\boldsymbol{\delta\sigma}^\lambda$, as shown in Fig.~\ref{fig:tank_prior_short}. In this experiment, the standard deviation of the initial reconstruction within region $D$ was $\operatorname{std}\!\left(\boldsymbol{\delta\sigma}^\lambda|_D\right)=0.000257$, and we set $\alpha=6.99$. We computed the kernel width parameter from Eq.~\eqref{eq:kernel_width} as $h=0.001794$. Then, we substituted $h$ and $\boldsymbol{\delta\sigma}^\lambda$ into Eq.~\eqref{eq:Gaussian_similarity_matrix} to construct the Gaussian similarity matrix $\mathbf{C}$. According to the minimum-standard-deviation criterion within region $D$ defined in Eq.~\eqref{eq:selected_eigenvector}, we selected the first principal eigenvector of matrix $\mathbf{C}$ and used the condition $\mathbf{v}_1<0.0010$ for threshold segmentation. The first four principal eigenvectors are shown in Fig.~\ref{fig:tank_eigs_first4_short}. We set the experimental parameters of the proposed algorithm as $\lambda=5.0\times10^{-8}$, $\rho=1$, $\lambda_1=1.0\times10^{12}$, $\lambda_2=3.0\times10^{12}$, and $\beta=3.0\times10^{-5}$; we set the maximum number of iterations to 300 and the stopping tolerance to $10^{-5}$.

\begin{figure}[htbp]
    \centering
    \begin{minipage}{0.24\columnwidth}
        \centering
        \includegraphics[width=\linewidth]{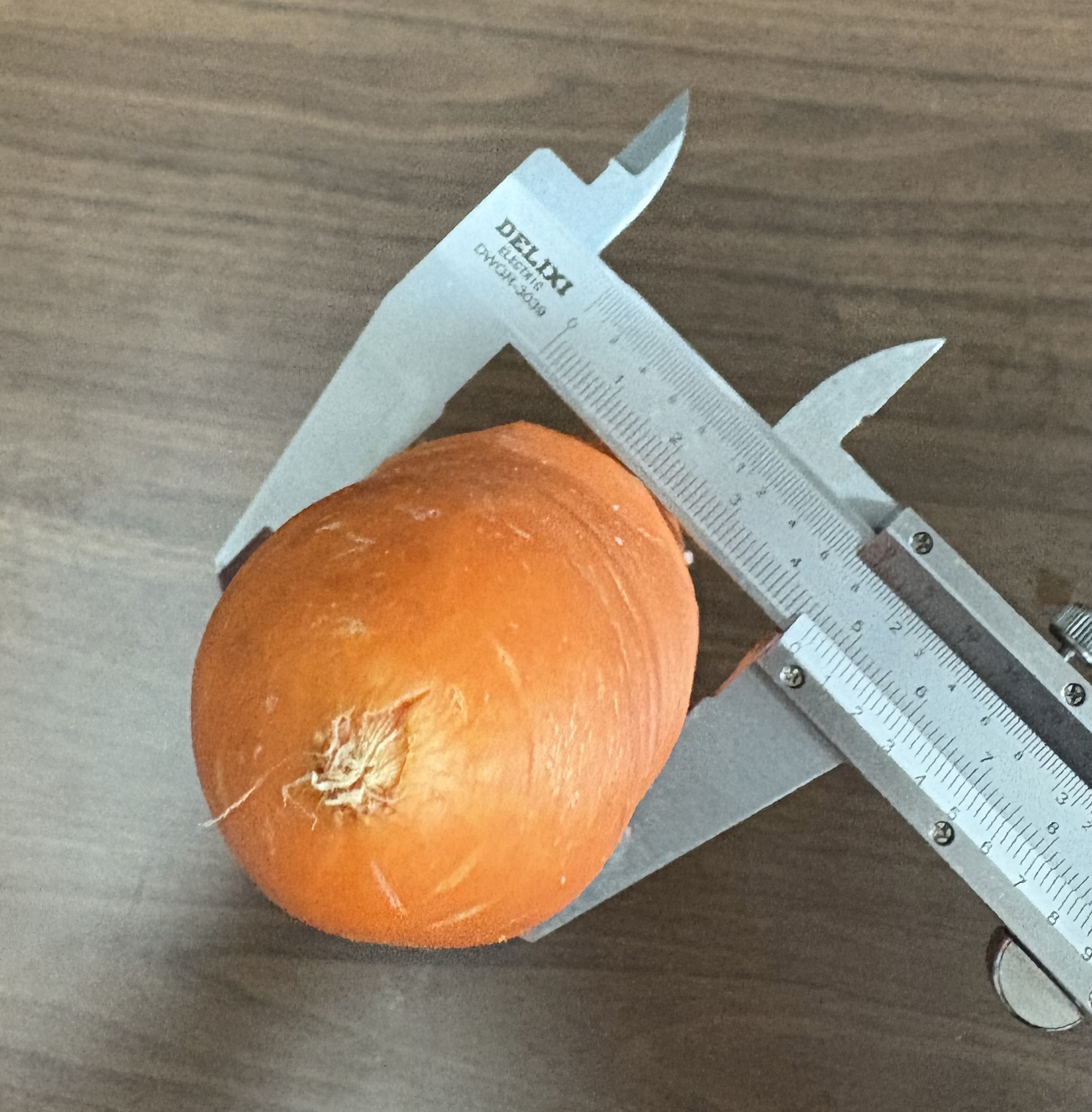}
        \centerline{\footnotesize (a)}
    \end{minipage}
    \hspace{0.02\columnwidth}
    \begin{minipage}{0.33\columnwidth}
        \centering
        \includegraphics[width=\linewidth]{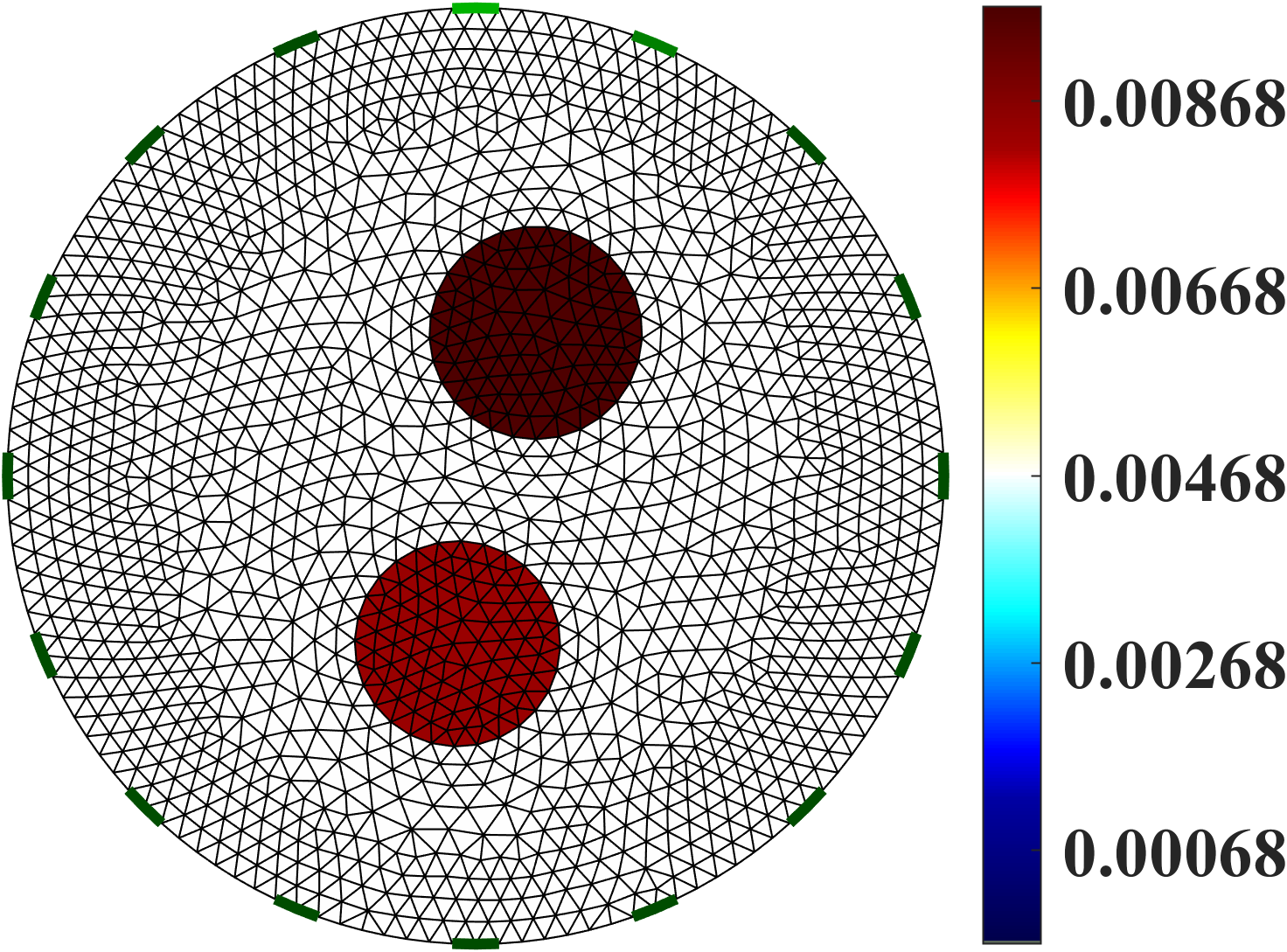}
        \centerline{\footnotesize (b)}
    \end{minipage}
    \hspace{0.005\columnwidth}
    \begin{minipage}{0.33\columnwidth}
        \centering
        \includegraphics[width=\linewidth]{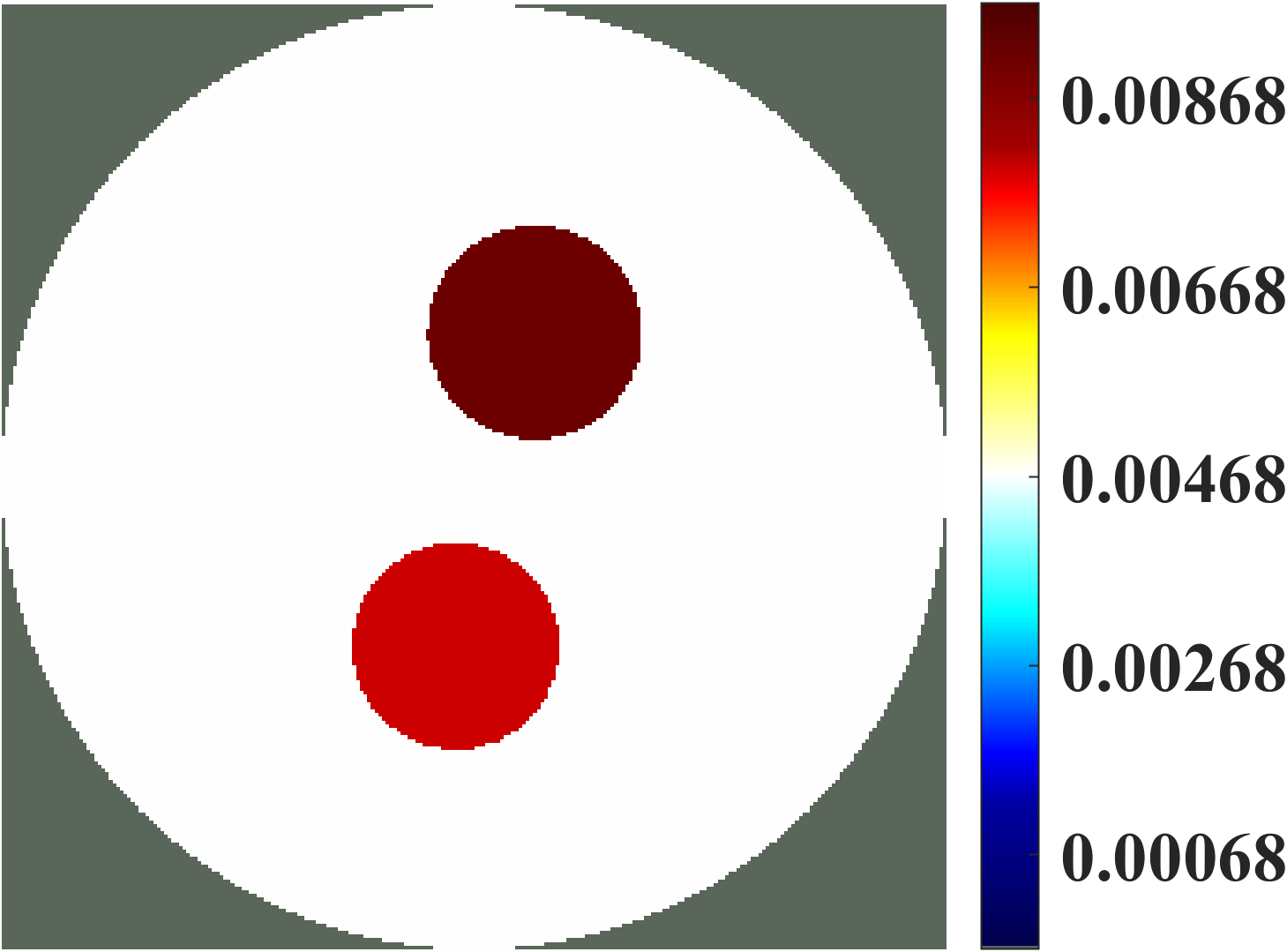}
        \centerline{\footnotesize (c)}
    \end{minipage}
    \caption{Construction process of the equivalent reference model for the carrot--cucumber tank experiment. (a) Example of vernier-caliper measurement at the marked immersion height. (b) High-resolution reference finite element model. (c) Corresponding $256\times256$ reference conductivity image.}
    \label{fig:tank_reference_short}
\end{figure}

In this experiment, we constructed an equivalent reference model for posterior quantitative evaluation. First, we approximated the cross sections of the carrot and cucumber on the imaging plane as circles and measured their diameters along multiple directions. After averaging, the diameters were $4.40\,\mathrm{cm}$ and $4.25\,\mathrm{cm}$, respectively. By normalizing them with the tank inner diameter of $19.35\,\mathrm{cm}$, we obtained the corresponding equivalent radii. Then, we performed connected-component separation on the foreground region where $\mathbf{M}=1$ and extracted the geometric centroids of the two connected subregions as the centers of the equivalent carrot and cucumber targets, respectively. Based on the original two-dimensional finite element reconstruction mesh and the above centers and equivalent radii, we regenerated a high-resolution reference finite element mesh \cite{adler2006eidors} and assigned conductivity values to the two target regions separately. For each target region, we assigned the mean conductivity obtained from the Isotropic TV reconstruction within the corresponding equivalent target region. This assignment strategy differs from that used in the `mean prior' $\widetilde{\boldsymbol{\delta\sigma}}$, where the entire foreground region $\mathcal{F}$ is assigned a single regional mean value. This reference mesh contains 2804 triangular elements and 1467 nodes, as shown in Fig.~\ref{fig:tank_reference_short}(b). Finally, we interpolated it onto a unified $256\times256$ image grid, as shown in Fig.~\ref{fig:tank_reference_short}(c). This equivalent reference model is used only for posterior quantitative evaluation; it does not represent the true conductivity distribution inside the real samples and does not participate in the reconstruction process.

\begin{figure*}[t]
    \centering
    \includegraphics[width=0.90\textwidth]{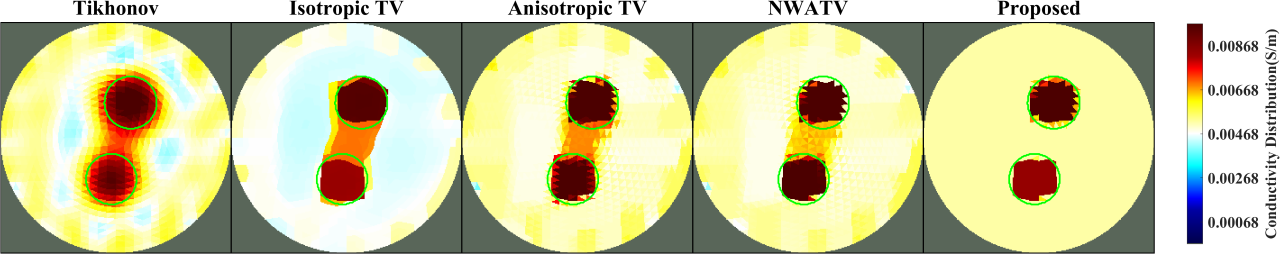}
    \caption{(a)-(e) illustrate the reconstruction results for the carrot-cucumber tank experiment using Tikhonov, Isotropic TV, Anisotropic TV, NWATV, and Proposed reconstructions, respectively. Green contours indicate the equivalent reference geometry.}
    \label{fig:tank_1x5}
\end{figure*}

Fig.~\ref{fig:tank_1x5} shows the reconstruction results using the Tikhonov, Isotropic TV, Anisotropic TV, NWATV, and the proposed method. It can be observed that the proposed method effectively suppresses diffused background artifacts while preserving target boundary sharpness, and can more stably distinguish adjacent targets.

\begin{figure}[htbp]
    \centering
    \begin{minipage}{0.48\columnwidth}
        \centering
        \includegraphics[width=\linewidth]{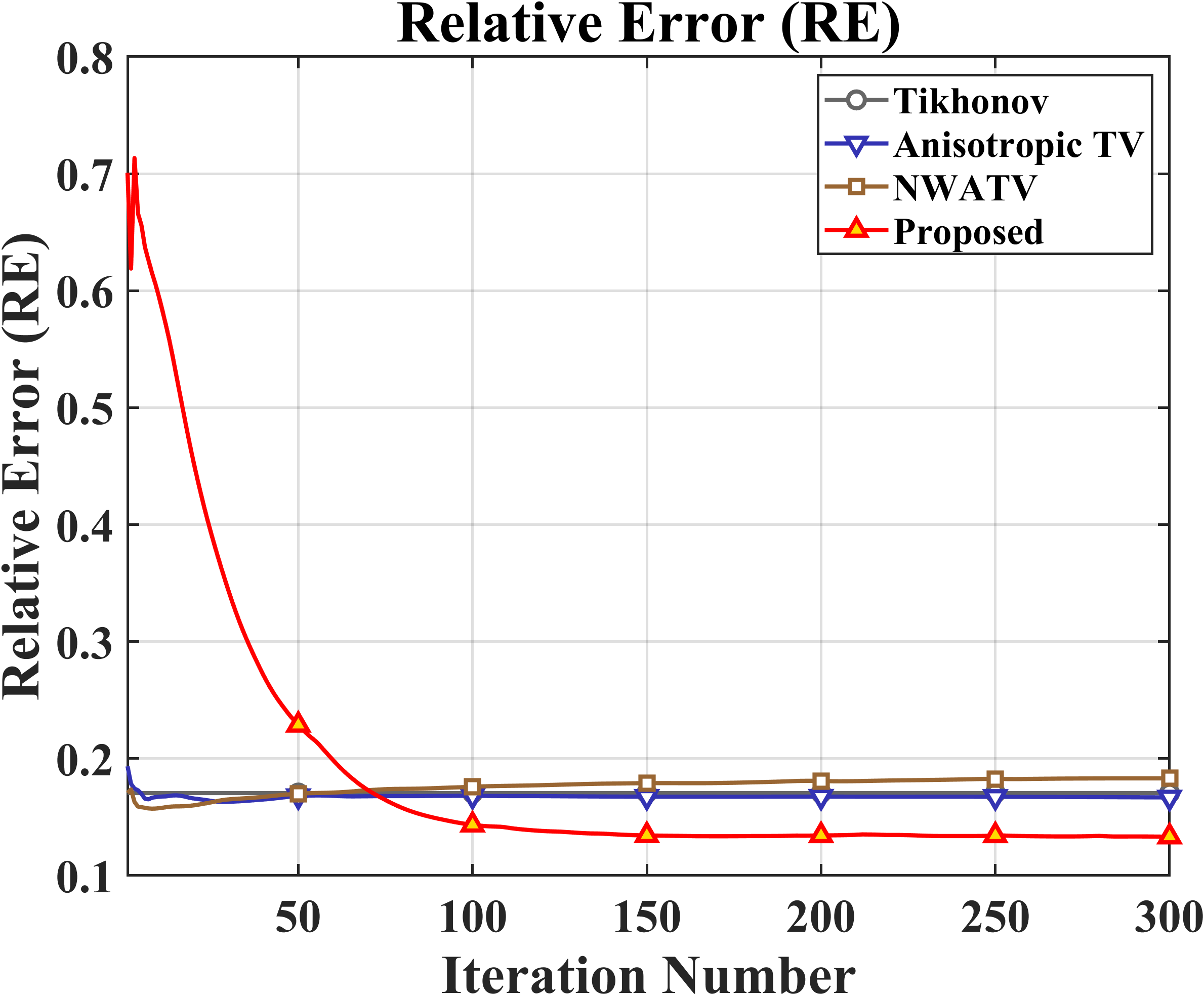}
        \centerline{\footnotesize (a) RE}
    \end{minipage}\hfill
    \begin{minipage}{0.48\columnwidth}
        \centering
        \includegraphics[width=\linewidth]{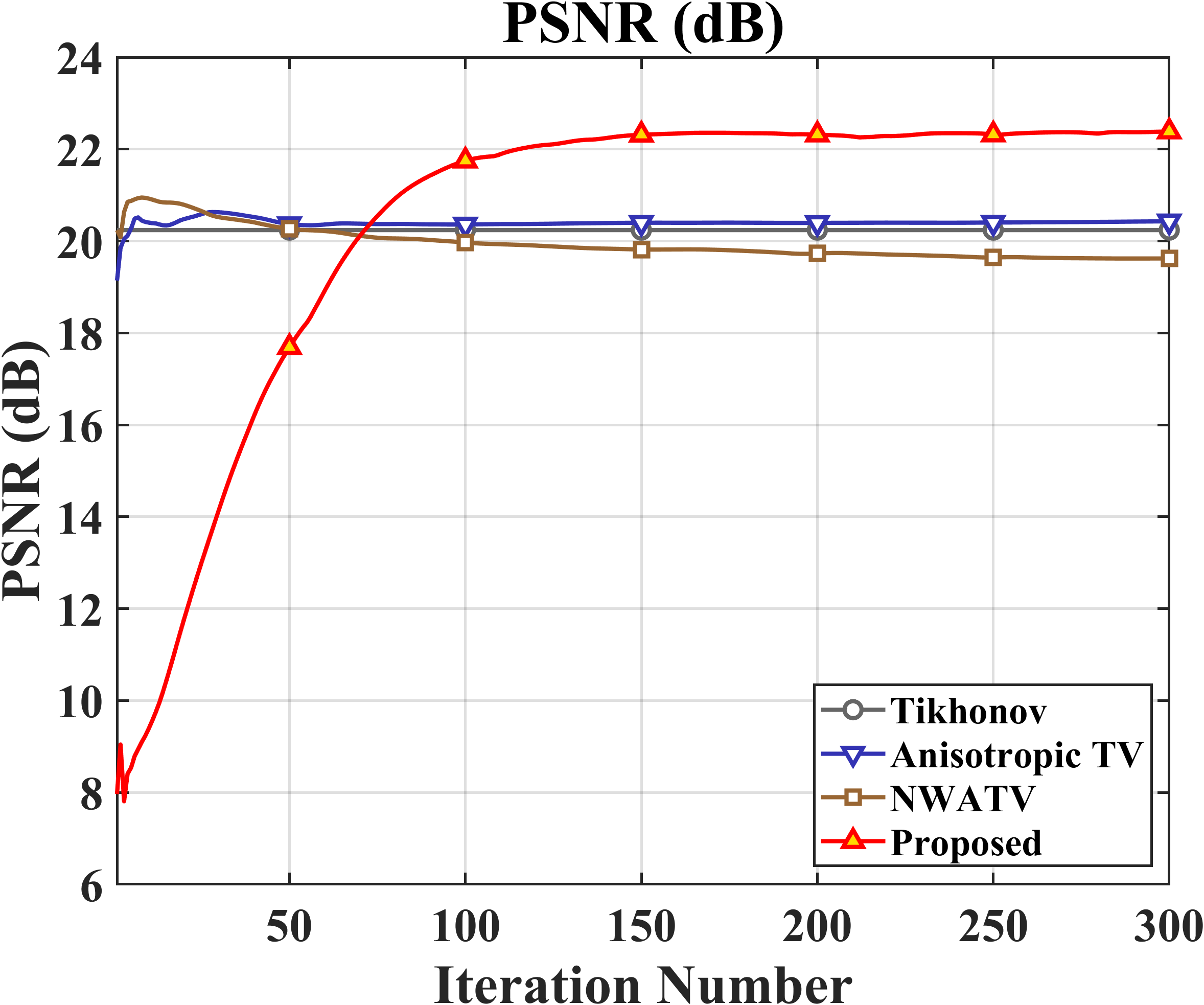}
        \centerline{\footnotesize (b) PSNR}
    \end{minipage}

    \vspace{0.2cm}

    \begin{minipage}{0.48\columnwidth}
        \centering
        \includegraphics[width=\linewidth]{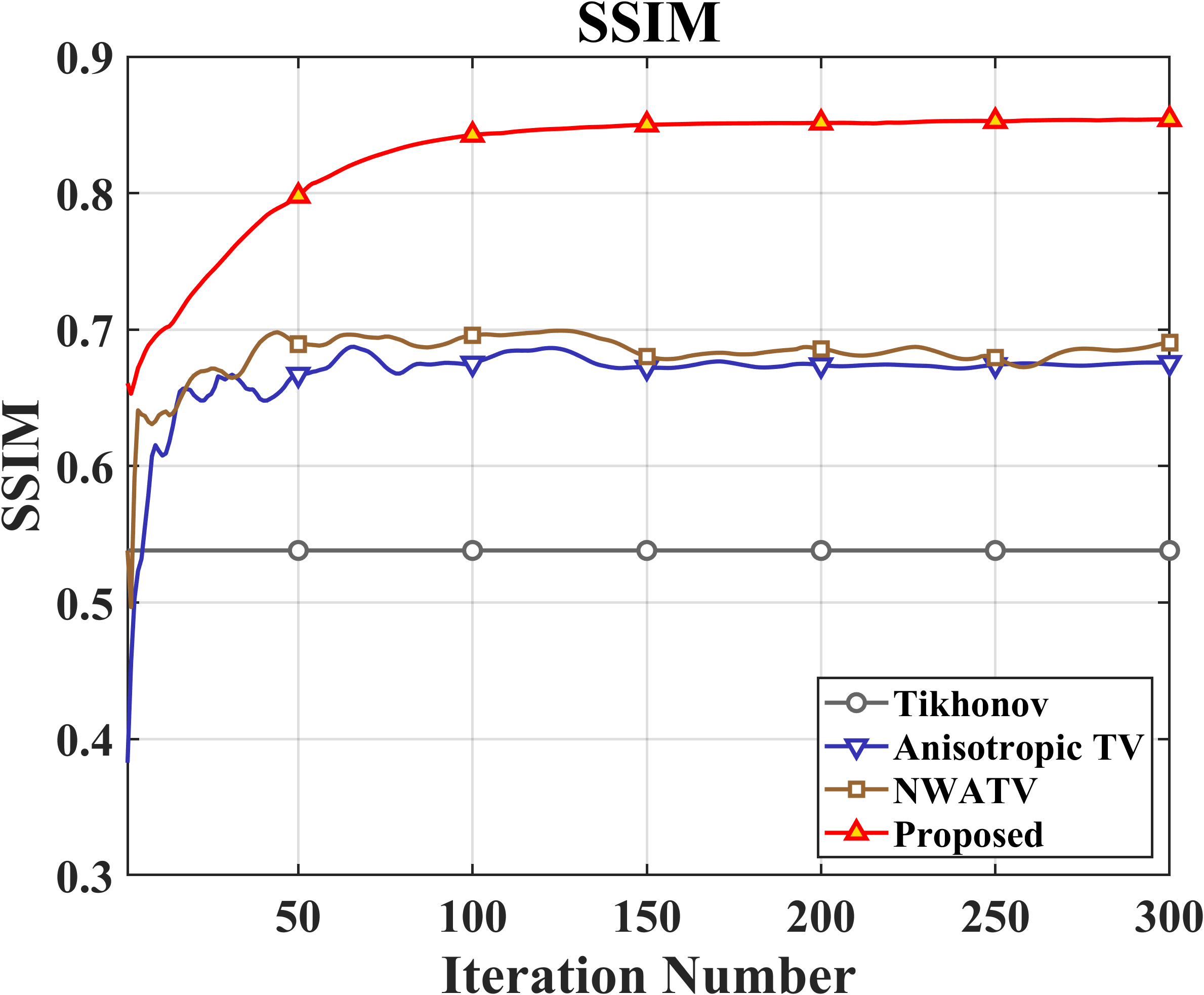}
        \centerline{\footnotesize (c) SSIM}
    \end{minipage}
    \caption{Comparison of the evolution curves of relative error (RE), peak signal-to-noise ratio (PSNR), and structural similarity index (SSIM) with respect to the iteration steps in the carrot--cucumber tank experiment.}
    \label{fig:tank_metrics_curves}
\end{figure}

Fig.~\ref{fig:tank_metrics_curves} shows the evolution curves of RE, PSNR, and SSIM in the carrot--cucumber tank experiment. Table~\ref{tab:tank_metrics} gives the corresponding final quantitative metrics. According to this table, the proposed method achieves the best results in RE, PSNR, and SSIM in this experiment.

\begin{table}[htbp]
    \centering
    \caption{Quantitative evaluation metrics for the carrot--cucumber tank experiment. Slash entries indicate that Isotropic TV was used to assign the equivalent reference amplitude and was therefore not numerically compared.}
    \label{tab:tank_metrics}
    \renewcommand{\arraystretch}{1.2}
    \begin{tabular}{lccc}
        \toprule
        \textbf{Algorithm} & \textbf{RE} & \textbf{PSNR (dB)} & \textbf{SSIM} \\
        \midrule
        Tikhonov & 0.1705 & 20.24 & 0.5382 \\
        Isotropic TV & / & / & / \\
        Anisotropic TV & 0.1668 & 20.43 & 0.6758 \\
        NWATV & 0.1831 & 19.62 & 0.6907 \\
        Proposed & \textbf{0.1332} & \textbf{22.38} & \textbf{0.8541} \\
        \bottomrule
    \end{tabular}
\end{table}

\section{Discussion}
\label{sec:discussion}

In this paper, we propose a local structural prior to address the target-merging problem caused by the spatially nonuniform sensitivity of EIT. For this, conventional methods use the CT image to generate the mask \cite{Ren2019}, while the proposed approach extracts a local structural prior from an initial reconstruction and incorporates it into a subsequent reconstruction. To be precise, we use an initial NWATV reconstruction as a reference image to construct a Gaussian similarity matrix. The eigendecomposition of this matrix provides an image-adaptive representation of the dominant organization of the reconstructed conductivity distribution. The selected eigenvector is then segmented to generate a structural mask, which is used to define a localized mean prior. Consequently, the proposed regularization together with a global edge-preserving regularity, the NWATV, is capable of addressing the target-merging problem in EIT.

Numerical results indicate that the proposed method improves the spatial separability of the two lung regions in the three-dimensional thorax model under both 50-dB and 30-dB noise conditions. The water tank experiment further demonstrates that the proposed method can be applied to measured EIT data. In the carrot-cucumber experiment, the proposed method could achieve an RE of 0.1332, a PSNR of 22.38~dB, and the highest SSIM of 0.8541. Compared with Tikhonov, anisotropic TV, and NWATV, the proposed method yielded a lower reconstruction error and higher PSNR and SSIM values.

In this section, we will validate the robustness of the reconstruction against the choice of $\alpha$, $D$. To this end, we use the three-dimensional numerical simulation results under the 50 dB noise condition as the analysis object.

To examine the sensitivity of the proposed method to the parameter $\alpha$, the originally selected fusion region is denoted as $D_0$, and the empirical parameter $\alpha=1.56$ is denoted as $\alpha_0$. The reference image is kept unchanged as the NWATV initial reconstruction $\delta{\bm \sigma}^\lambda$. With the region $D_0$ kept unchanged, two values, $\alpha=1$ and $\alpha=2$, are used for testing. The kernel width parameter $h$ is estimated according to Eq.\eqref{eq:kernel_width}, the Gaussian similarity matrix $\mathbf{C}$ is computed, and the mask $\mathbf{M}$ and the prior term $\widetilde{\delta{\bm \sigma}}$ are constructed based on $\mathbf{C}$. Under the condition that the reconstruction parameters are kept consistent with the baseline case $D_0,\alpha_0$, Eq.\eqref{minimization:final_regularized} is solved to obtain the reconstruction results of the proposed method. The results are shown in Table~\ref{tab:alpha_sensitivity}.

\begin{table}[htbp]
    \centering
    \caption{$\alpha$ sensitivity analysis results.}
    \label{tab:alpha_sensitivity}
    \renewcommand{\arraystretch}{1.2}
    \scriptsize
    \begin{tabular}{@{}lcccccc@{}}
        \toprule
        \textbf{$D$ type} & \textbf{$\alpha$} & \textbf{$\operatorname{std}(\delta{\bm \sigma}^\lambda|_{D})$} & \textbf{$h$} &  \textbf{$\mathbf{v}_1>\mbox{thresholder}$ } & \textbf{RE} \\
        \midrule
        $D_0$ & 1.56 & 0.019814 & 0.0310  & $>0.010$ & 0.061354 \\
        $D_0$ & 1.00 & 0.019814 & 0.0198  & $>0.002$ & 0.064098 \\
        $D_0$ & 2.00 & 0.019814 & 0.0396  & $>0.020$ & 0.061227 \\
        \bottomrule
    \end{tabular}
\end{table}

\begin{figure}[!t]
    \centering
    \begin{minipage}{0.35\columnwidth}
        \centering
        \includegraphics[width=\linewidth]{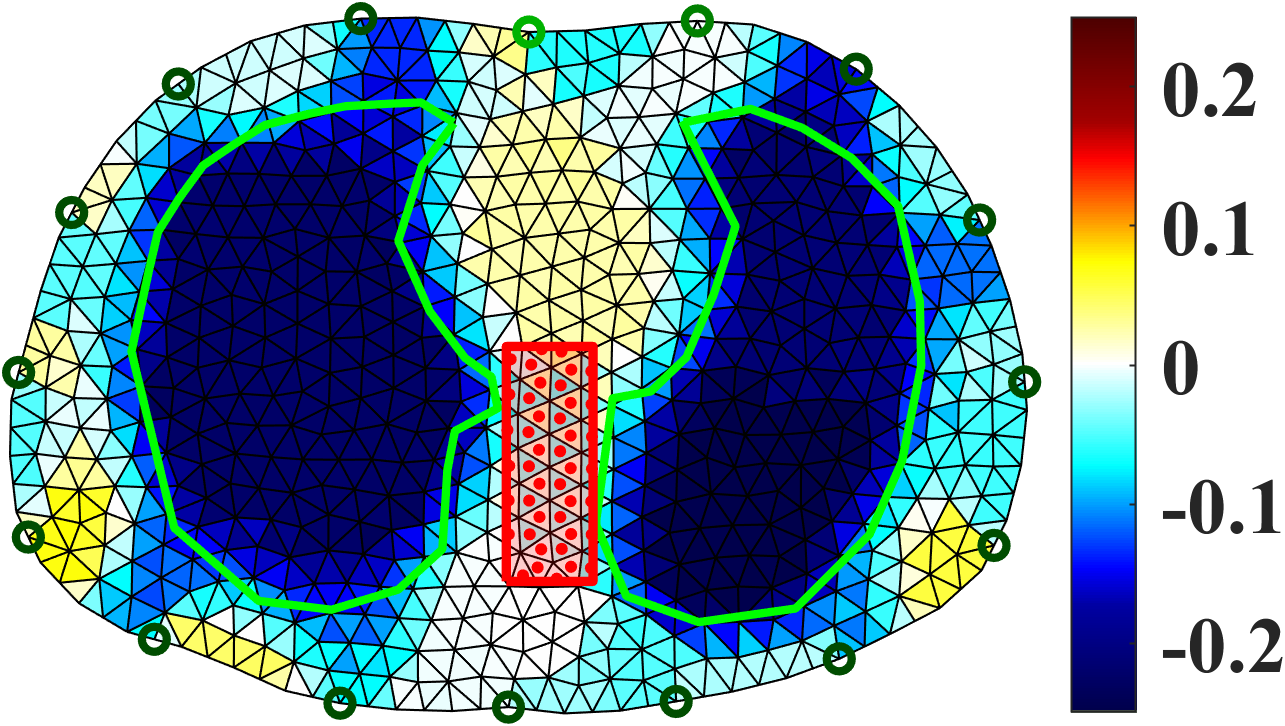}
        \centerline{\footnotesize (a)}
    \end{minipage}
    \hspace{0.06\columnwidth}
    \begin{minipage}{0.35\columnwidth}
        \centering
        \includegraphics[width=\linewidth]{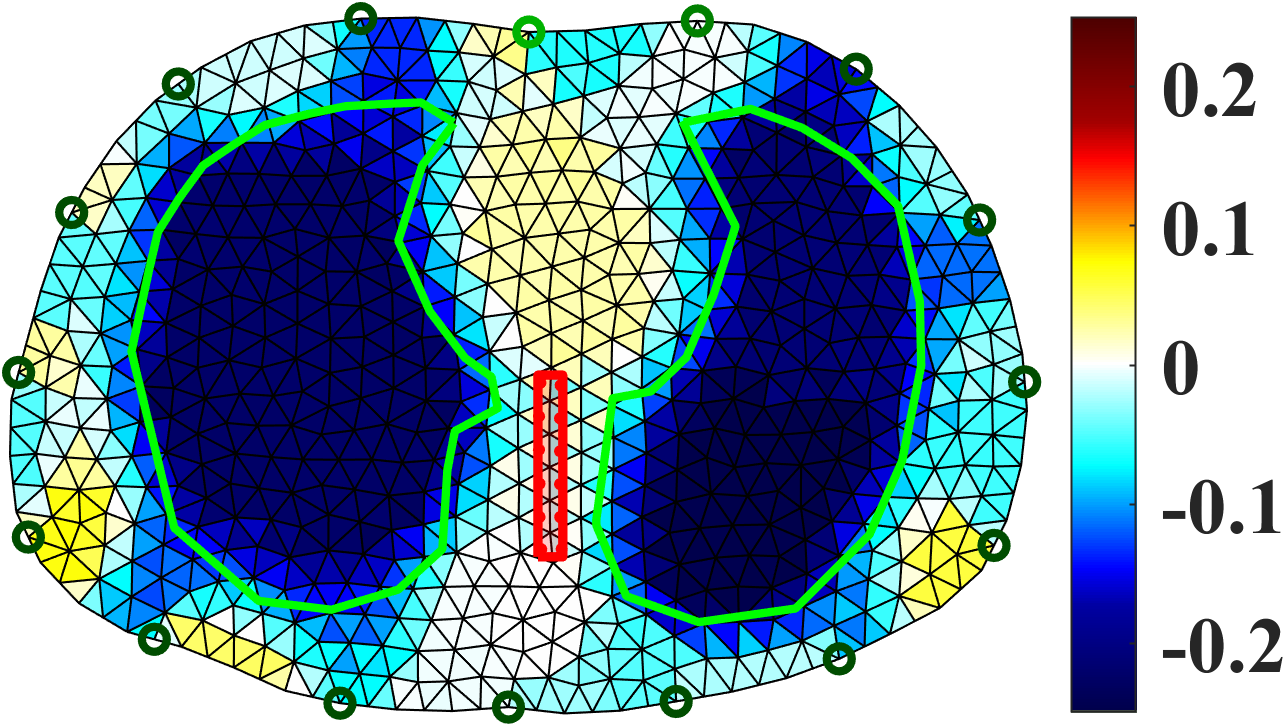}
        \centerline{\footnotesize (b)}
    \end{minipage}
    \caption{Illustration of the morphological variations of the fusion region $D$. (a) $D_1$ obtained by dilating $D_0$ by one layer. (b) $D_2$ obtained by eroding $D_0$ by one layer.}
    \label{fig:D_sensitivity_roi}
\end{figure}

Furthermore, to analyze the influence of the selection of the fusion region $D$ on the proposed method, we obtain $D_1$ and $D_2$ by applying morphological dilation and erosion to the originally selected fusion region $D_0$ by one layer, respectively. The corresponding region selections are shown in Fig.~\ref{fig:D_sensitivity_roi}. For $D_1$ and $D_2$, $\operatorname{std}(\delta{\bm \sigma}^\lambda|_{D_1})$ and $\operatorname{std}(\delta{\bm \sigma}^\lambda|_{D_2})$ are calculated, and a fixed coefficient $\alpha=1.5$, which is close to $\alpha_0=1.56$, is adopted. Then, the same procedure as in the previous paragraph is followed to calculate $\mathbf{C}$, $h$ and reconstruct $\mathbf{M}$, and $\widetilde{\delta{\bm \sigma}}$. Finally, reconstruction is performed while keeping the parameters in Eq.\eqref{minimization:final_regularized}. The results are shown in Table~\ref{tab:D_sensitivity}.

\begin{table}[!t]
    \centering
    \caption{Reconstruction results under different definitions of region $D$.}
    \label{tab:D_sensitivity}
    \renewcommand{\arraystretch}{1.2}
    \scriptsize
    \begin{tabular}{@{}lcccccc@{}}
        \toprule
        \textbf{$D$ type} & \textbf{$\alpha$} & \textbf{$\operatorname{std}(\delta{\bm \sigma}^{(0)}|_{D})$} & \textbf{$h$} & \textbf{$\mathbf{v}_1>\mbox{thresholder}$ } & \textbf{RE} \\
        \midrule
        $D_0$ & 1.56 & 0.019814 & 0.0310 &  $>0.010$ & 0.061354 \\
        $D_1$ & 1.50 & 0.020574 & 0.0309 &  $>0.010$ & 0.061227 \\
        $D_2$ & 1.50 & 0.021466 & 0.0322 &  $>0.010$ & 0.061763 \\
        \bottomrule
    \end{tabular}
\end{table}

As shown in Table~\ref{tab:alpha_sensitivity}, when $\alpha$ increases from 1 to 2, the corresponding range of $h$ is 0.0198--0.0396. However, the RE values of the three reconstruction results differ only slightly and are all lower than those of the other comparison methods, indicating that the proposed method is insensitive to variations in $\alpha$ within this specific range. As shown in Table~\ref{tab:D_sensitivity}, when the fusion region is dilated or eroded by one layer based on $D_0$, the estimated values of $h$ remain close to that in the $D_0$ case, and the eigenvector selected by Eq.\eqref{eq:selected_eigenvector} is consistently $k^\ast=1$. The RE values of the three results differ only slightly and are all lower than those of the other comparison methods. This phenomenon demonstrates that the proposed method is stable with respect to moderate morphological variations of the fusion region $D$. It should be noted that although the $D_1$ ($\alpha = 1.5$) case and the $D_0\; (\alpha=2)$ case use different values of $h$ and different truncation thresholds, the mask $\mathbf{M}$ and the prior term$\widetilde{\delta{\bm \sigma}}$ obtained after truncation are exactly the same, and therefore the two cases finally yield the same RE.

\section{Conclusion and future works}
\label{sec:conclusion}

In this work, we proposed a self-generated local structural prior for reducing target-merging artifacts in electrical impedance tomography. The proposed method first obtains an initial conductivity reconstruction using NWATV regularization. A Gaussian similarity matrix is then constructed from the intensity similarities between image elements. An eigenvector is adaptively selected according to its stability within a prescribed region, and the selected eigenvector is segmented to generate a localized structural mask. This mask is further used to construct a mean-prior conductivity distribution and to impose local structural constraints in a subsequent reconstruction.

The proposed strategy has two main advantages. First, the structural prior is generated directly from the EIT reconstruction itself and therefore does not require additional anatomical imaging data or a pretrained network. Second, the prior is applied locally, allowing the method to improve the separation of neighboring targets without imposing a uniform constraint over the entire imaging domain.

The three-dimensional numerical experiments demonstrated that the proposed method preserved the separation between the two lung regions under both 50-dB and 30-dB measurement-noise conditions. The water-tank experiment with carrot and cucumber targets further confirmed the applicability of the method to measured EIT data. These results indicate that a self-generated local structural prior can effectively complement global anisotropic total-variation regularization and improve the spatial separability of neighboring conductivity targets.

Future work will focus on developing automatic strategies for selecting the problematic region, kernel width, and segmentation threshold; automatic regularization parameter selection; and validating the method using human lung imaging experiments.

\end{document}